\documentclass[%
superscriptaddress,
  aps,
 prx,
 twocolumn
]{revtex4-2}
\usepackage{amsmath}
\usepackage{natbib}
\usepackage{amsfonts}
\usepackage{amssymb}
\usepackage{dsfont}
\usepackage{multirow}
\usepackage[normalem]{ulem}
\usepackage[caption=false]{subfig}
\usepackage{graphicx}
\usepackage[export]{adjustbox}
\usepackage{enumitem}
\usepackage{color}
\usepackage{bbold}
\usepackage{cancel}
\usepackage{comment}

\allowdisplaybreaks[4]

\newcommand{\T}{\mathrm{T}}
\newcommand{\E}{\mathrm{E}}

\newcommand{\C}{{\mathrm{C}}}
\newcommand{\U}{{\mathrm{U}}}
\newcommand{\SC}{{\mathrm{S}}}
\newcommand{\X}{{\mathrm{X}}}
\newcommand{\M}{{\mathcal{M}}}
\newcommand{\fd}{{\vphantom{\dagger}}}
\newcommand{\sh}{\,\mathrm{sh}\,}
\newcommand{\ch}{\,\mathrm{ch}\,}
\newcommand{\im}{\mathrm{i}}

\makeatletter
\DeclareRobustCommand{\rvec}[1]{%
  \mathpalette\do@rvec{#1}%
}
\newcommand{\do@rvec}[2]{%
  \fix@rvec{#1}{+}%
  \reflectbox{$\m@th#1\vec{\reflectbox{$\fix@rvec{#1}{-}\m@th#1#2\fix@rvec{#1}{+}$}}$}%
  \fix@rvec{#1}{-}%
}
\newcommand{\fix@rvec}[2]{%
  \ifx#1\displaystyle
    \mkern#23mu
  \else
    \ifx#1\textstyle
      \mkern#23mu
    \else
      \ifx#1\scriptstyle
        \mkern#22mu
      \else
        \mkern#22mu
      \fi
    \fi
  \fi
}
\makeatother

\makeatletter
\newcommand{\lowersim}[2]{%
  \sbox\z@{$#1<$}%
  \raisebox{-\dimexpr\height-\ht\z@}{$\m@th#1#2$}%
}
\makeatother

\begin{document}
\title{Symmetry classes of open fermionic quantum matter}

\author{Alexander Altland}
\affiliation{Institut f\"{u}r Theoretische Physik, Universit\"{a}t zu K\"{o}ln, D-50937 Cologne, Germany}

\author{Michael Fleischhauer}
\affiliation{Department of Physics and Research Center OPTIMAS, University of Kaiserslautern, 67663 Kaiserslautern, Germany}

\author{Sebastian Diehl}
\affiliation{Institut f\"{u}r Theoretische Physik, Universit\"{a}t zu K\"{o}ln, D-50937 Cologne, Germany}

\begin{abstract}
We present a full symmetry classification  of fermion matter in and out of thermal equilibrium.
Our approach starts from first principles, the ten different classes of linear and
anti-linear state transformations in fermionic Fock spaces, and symmetries defined
via invariance properties of the dynamical equation for the density matrix. The
object of classification are then the generators of reversible dynamics, dissipation
and fluctuations featuring in the generally irreversible and interacting dynamical
equations. A sharp distinction between the symmetries of equilibrium and out of equilibrium dynamics, respectively, arises from the different role played by `time' in these two cases: In unitary quantum
mechanics as well as in  `micro-reversible'  thermal equilibrium, 
anti-linear transformations  combined with an inversion of 
time define time reversal symmetry. However, out of equilibrium an inversion of time becomes
meaningless, while  anti--linear transformations in Fock space remain physically significant, and 
hence must be considered in autonomy. The practical consequence of this dichotomy is a  novel realization of antilinear symmetries (six out of the ten fundamental classes) in non-equilibrium quantum dynamics that is fundamentally different from the established rules of thermal equilibrium. At large times, the dynamical generators thus symmetry classified  determine the  steady state non-equilibrium distributions for arbitrary
interacting systems. To illustrate this principle, we consider the fixation of a symmetry protected topological phase in a system of interacting lattice fermions. More generally, we consider the practically important class of mean field interacting systems, represented by Gaussian states. This class is naturally described in the language of  non-Hermitian matrices, which allows us to compare to previous
classification schemes in the literature.

\end{abstract}

\date{\today}
\maketitle

%
%
%

\section{Introduction} 
\label{sec:introduction}

The distinction between different   unitary and anti-unitary
symmetries~\cite{Wigner51,Wigner58,Dyson62,Dyson62_2} is a powerful organizing principle in the classification of quantum
matter. It has been spectacularly successful in the description of gapped fermionic
matter, where the identification of topologically twisted ground states
 on the background of ten fundamental symmetry classes  \cite{Altland97} culminated in the periodic table of
topological insulators and superconductors~\cite{Ryu10,Kitaev09}.

In this paper we ask  how the concept of symmetry classifications can be generalized
to fermion matter  pushed out of thermal equilibrium by an external  environment.
This question is motivated in part by recent experimental progress in the physics of
condensed matter, atomic condensates, and optics, which led to the realization of
novel phases of quantum matter in engineered
environments~\cite{Zeuner15,Zhou18,Weimann17,Xiao20,Helbig2020,Hodaei17,Chen17,cerjan19,Poli15}.
These developments call for the  classification of symmetries and topologies of open
quantum matter, in extension of existing frameworks for closed system quantum ground
states. 

Earlier work in this direction has put the emphasis on the most apparent
consequence of environmental coupling, lossy dynamics and its description in terms of non-Hermitian
matrix operators~\cite{Leclair02,Ueda19,Lee19,ashida2020nonHermitian,huber2020emergence}. 
We here take a more general perspective and note that a comprehensive description of out of equilibrium symmetries must account for the interplay of dissipation and \emph{environmental fluctuations}. Our approach to the full problem  starts with the realization just how straightforward the description of
symmetries in the complementary case of isolated systems actually is.
There, the full information is stored in the symmetries of a single Hermitian
operator, the system Hamiltonian, $\hat H$. The latter encodes the symmetries on the
microscopic level via the definition of $\hat H$ from Fock space operators, it
describes the symmetries of state evolution via the evolution operator $\hat
U=\exp(- \im\hat H t)$ ($\hbar =1$), and those of long time stationary states through projectors
onto the many body ground state of $\hat H$, or, slightly more generally, a thermal
distribution $\hat\rho\sim\exp(-\hat H/T)$.

Out of equilibrium, the situation becomes distinctly more complex. How does an exhaustive set of symmetries describing a system out of equilibrium look like?  Which elements of the theory assume the role of the Hamiltonian in the description of these symmetries? And on the basis of what physical principles should they be described in mathematical terms? The formulation of concrete, and surprisingly simple answers to these questions is the mission of the research reported in this paper. 

\begin{figure}
\centering
\includegraphics[width=8.5cm]{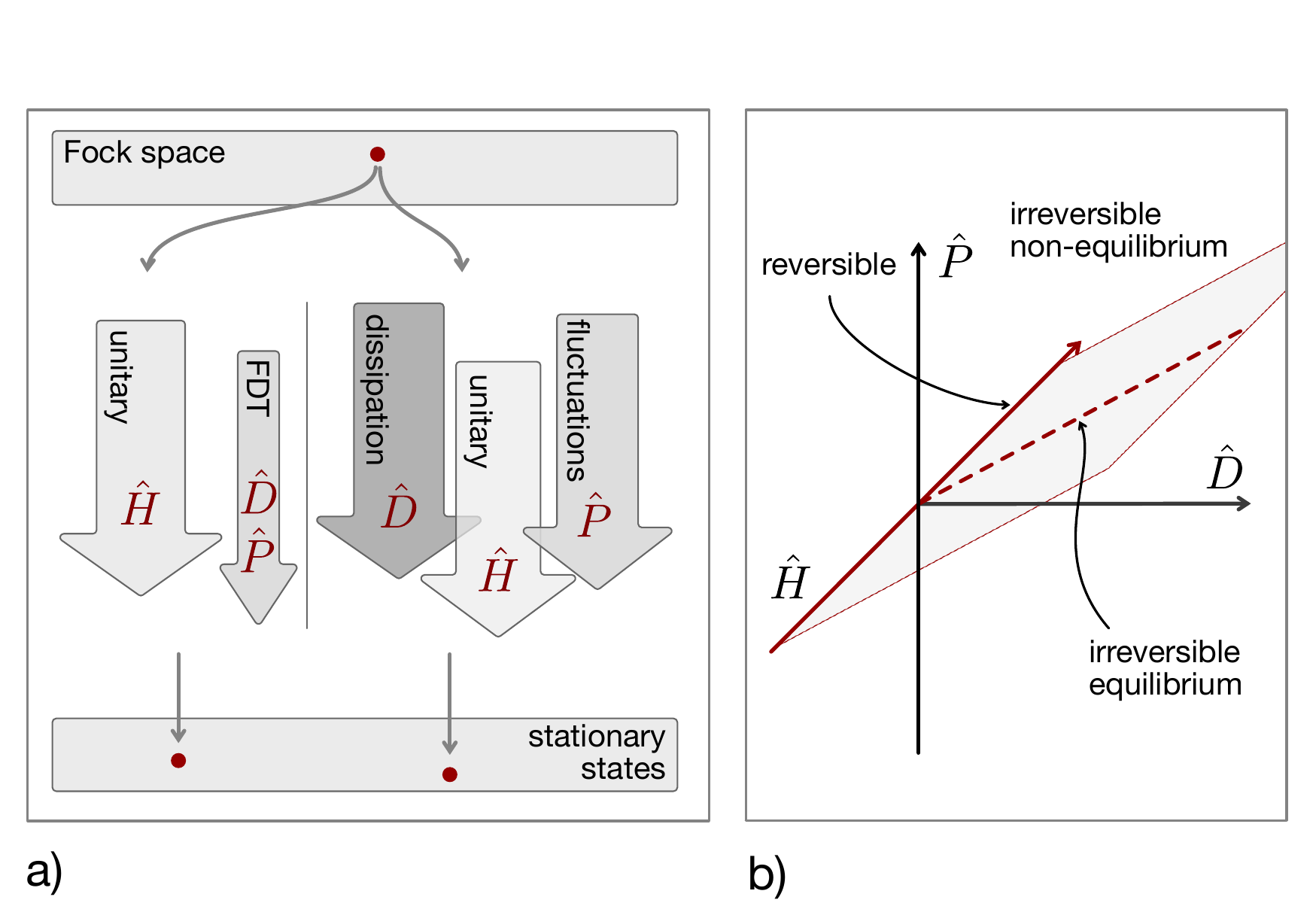}
\caption{\label{fig1}  
Building blocks entering the description of state evolution of fermionic quantum
 matter. a) The evolution of Fock space quantum states primarily depends on whether a system is in or out of equilibrium. In equilibrium, left, both the dynamics and the stationary state are controlled by a Hamiltonian, $\hat H$. (Thermalization at a given ambient temperature, $T$, may be effected by fluctuation and dissipation, subject to the fluctuation-dissipation theorem (FDT), Eq.~\eqref{eq:FDT}.)
 Out of equilibrium, the dynamics is specified in terms of three generators, $\hat H,
 \hat D, \hat P$, representing unitary evolution, dissipation, and fluctuations,
 respectively. In either case, initial Fock space states evolve into stationary states whose symmetries are determined by the dynamical generators, but not the initial configurations.  b) In equilibrium dynamics, fluctuation and dissipation are locked
 by the fluctuation-dissipation relation, Eq.~\eqref{eq:FDT}, reducing the `operator
 space' to two freedoms, $\hat H$ and $\hat D$ controlling the relaxation into an
 equilibrium configuration.   Anti-unitary symmetries appear in combination with an
 inversion of time. The representation of anti-unitary symmetries, $\T$, depends on
 whether the dynamics is unitary $\hat D=\hat P=0$ ($\T$ in combination with time
 reversal, $t\to -t$), irreversible equilibrium $\hat D\propto \hat P$ ($\T$ in
 combination with thermal time reversal, $t\to -t +\im \beta$), or irreversible out of
 equilibrium ($\hat T$ in autonomy).}
\end{figure}

To get warmed up to the subject, consider a system whose dynamics is subject to
damping, external driving, and an intrinsic Hamiltonian. In its evolution, these
influences manifest themselves in the competition of dissipation, fluctuations, and
unitary dynamics (cf. Fig.~\ref{fig1}). We aim to understand in which ways these
three are constrained by symmetries.  On top of the concrete  challenges formulated
above, we are immediately facing a fundamental issue, the status of \emph{time
reversal}. Six out of ten of the fundamental symmetries of equilibrium quantum matter
make reference to this symmetry. However, out of equilibrium, time reversal looses
its physical meaning. Does this mean that the number of symmetries reduces from ten
to four? The answer is no. To understand what is happening, recall that time reversal
in quantum mechanics is implemented by a combined operation inverting time, $t\to
-t$, and subjecting operators to an anti-unitary transformation. Out of equilibrium,
the former looses its meaning, but the latter does not. We are thus led to investigate
the status of anti-linear operations in autonomy. 

This realization, surprisingly, appears to be novel, and it leads to
unexpected structures. To mention one example, we call a quantum Hamiltonian `chiral', 
stabilizing a chiral equilibrium phase,
if it anti-commutes with a Pauli matrix $\sigma_3$ in some representation. However,
it turns out that a Hamiltonian participating in the stabilization of a chiral out
of equilibrium phase must \emph{commute} with $\sigma_3$. Such constraints have practical bearings, for example for the engineered preparation of symmetry enriched out of equilibrium phases as we will show. 

Returning to the general topic, our discussion below  thus focuses on the manifestations
of symmetries in the dynamical evolution of quantum states. The approach
starts from first principles with a representation of states via density operators
$\hat \rho(\{a_i,a_i^\dagger\})$ in Fock spaces spanned by creation and annihilation
operators $\{a_i,a_i^\dag\}$. Symmetry operations $\X$ are realized in ten families
of unitary and anti-unitary transformations of these operators, nowadays mostly
labeled by the Cartan symbols, A, AIII, AI,$\ldots $, CI (see
Appendix~\ref{sec:SymmetryClassesTable} for a review). These symmetries are
well documented, meaning that the microscopically formulated starting point of the
theory is under control. In particular, the ten-fold classification implies the existence of a  unique symmetry label assigned to individual quantum
states at each instance of time. For example, in the  case of thermodynamic equilibrium,  $\hat \rho_\text{eq}
(\{a_i,a_i^\dag\})= e^{-\beta \hat H (\{a_i,a_i^\dag\})}$, the state inherits the symmetries of the  Hamiltonian $\hat H$. More generally, however, the classification statement remains formal before the state is described in the context of its \emph{dynamical evolution}.

The dynamics of quantum states is described by  evolution equations of the structure
\begin{align}
  \label{eq:MarkovianEvolutionGeneral}
  \partial_t \hat \rho(t)=\hat{\mathcal{X}}\hat
\rho(t).
\end{align}
This equation describes both Markovian state evolution, where $\hat{\mathcal{X}}\hat \rho(t)=\hat{\mathcal{X}}(t)\hat \rho(t)$ is time local  and non-Markovian cases where $\hat{\mathcal{X}}\hat
\rho(t)=\int_\Bbb{R^+} ds\, \hat{\mathcal{X}}(s)\hat\rho(t-s)$ contains a retarded convolution over time. In either case, the symmetries of the \emph{linear dynamical generator}
$\hat{\mathcal{X}}$ and their manifestations in the solutions $\hat \rho$ are the central topic of this work. Our discussion will be general in that it treats interacting and non-interacting systems on the same footing.

We first note that Fock space symmetry
transformations $\mathrm X$ affect the operators $\hat \rho(\{a_i,a_i^\dag\})$ and $
\hat{\mathcal{X}}(\{a_i,a_i^\dag\})$ featuring in the equation as $\hat \rho \to \hat
\rho_{\mathrm X}$ and $\hat{\mathcal{X}}\to \hat{\mathcal{X}}_\mathrm{X}$ via  their
representation on the creation and annihilation operators. Importantly, the
realization of a symmetry in the evolution equation may include an additional
transformation of the dynamical parameter `time', $t$, itself. A transformation is called
a \emph{symmetry}, if it leaves the evolution equation invariant, i.e. if the equation looks the same before and after the transformation. If this condition is met, $\hat \rho$ and $\hat \rho_\mathrm{X}$ solve the same equation and are therefore identical; the symmetry of the dynamically evolving state is established, including in the limit $\hat \rho(t\to\infty)$, where it defines the stationary phases. 

The moment, the dynamics includes elements of irreversibility induced by the coupling to an environment, the inversion of time $t\to -t$ in Eq.~\eqref{eq:MarkovianEvolutionGeneral} is no longer physical ---  movies of irreversible processes  don't make sense in reverse. The generic situation in this case is that of non-equilibrium dynamics, which will also be the focus of this paper. With inversion of time  
 out of the picture, the invariance condition reads  $ \hat{\mathcal{X}}_\mathrm{X}=+  \hat{\mathcal{X}}$. More specifically, the
 generators, $\hat{\mathcal{X}}=\hat{\mathcal{X}}(\hat H, \hat
D,\hat P)$ can generally be expressed in terms of three subordinate operators,
describing the contributions of unitary evolution, dissipation and fluctuations to
the dynamics. In this representation,  the symmetry criterion splits into three, individually for these operators. 

The symmetry classification of the generator of \emph{dynamics} $\hat{\mathcal{X}}$ defines the criteria required for the stabilization  of a (stationary) \emph{state} of definite symmetry.  We already mentioned that the  criteria obtained in this way differ from those of equilibrium systems. For example, the conditions for the Hamiltonian contribution to a dynamical evolution assume a form opposite to those in the equilibrium case. Given the scarcity of general principles characterizing out of equilibrium quantum distributions, the specification of symmetry criteria universally described in terms of the generators  $\hat H, \hat D,\hat P$ is an important contribution of this work.

At this point, we have mentioned two settings, the limit of closed system unitary
dynamics, and that of non-equilibrium irreversible dynamics. However, sandwiched
between these two, we have a third major 
class, that of \emph{irreversible equilibrium dynamics}, commonly associated to the
physics of thermalization. Thermalization is irreversible, and as in the
non-equilibrium case, a naive inversion of time in
Eq.~\eqref{eq:MarkovianEvolutionGeneral} is not physical. However, unlike in the
non-equilibrium case,  there still applies a principle of `micro-reversibility'. In
essence, it states that  the rates of microscopic processes are determined by  those
of their time reversed inverse processes. Micro-reversibility implies a symmetry
under shift inversion $\Delta t\to -\Delta t + \im \beta$, where $\beta$ is inverse
temperature, and $\Delta t$ the \emph{difference} entering the correlation of
observables at different times. In order for this condition to hold, fluctuations,
$P$, and dissipation, $D$ must be locked to each other via the fluctuation
dissipation theorem Eq.~\eqref{eq:FDT} below. From the larger perspective of out of
equilibrium dynamics, thermal equilibrium thus defines a `fine tuned' case,
much as unitary dynamics ($D=P=0$) is an even stronger confined limit. Either limit
comes with symmetry principles specific to its constraints, and different from the
general case. The  hierarchy of different settings is illustrated in Fig.~\ref{fig1}
b).

\subsection{Synopsis and summary of results} 
\label{sub:synopsis_and_summary_of_results}

We now turn to a more concrete level and summarize the main findings of our work. Proceeding in a bottom up manner, we first discuss the realization of symmetries in fermionic Fock space, before turning to their representation in different descriptions of effective dynamics. We also compare our results to related work on symmetry classification of open quantum systems. 

\noindent \emph{Symmetries --} 
Our starting point is the representation of
symmetries as unitary or anti-unitary transformations  in Fock space through their
action on  fermion operators $\{a_i^\fd,a_i^\dagger\}$. These operators represent the system after tracing over the
environmental degrees of freedom, and no reference
to a particular type of dynamics is made yet. Following the reasoning of Refs.~\cite{Altland97,Heinzner05}, all we
can say at this level is that modulo unitary equivalence ten different classes of transformations need to be
distinguished.  

More specifically, the basic operations from which these symmetries are
built by composition are an anti-unitary transformation ($\T \, \im\, \T^{-1}=-\im$) acting as  $\T a_i \T^{-1}=u_{\T,ij} a_j 
$, where $u_\T$
is a unitary matrix, and a unitary operation exchanging annihilators and creators $\C
a^\fd_j\C^{-1}=u_{\C,ij}a_j^\dagger$ (see Eq.~\eqref{eq:UTCSecondQuantized} for a
more detailed representation).
For bookkeeping purposes, we also define the  combination $\SC=\T\circ\C$
as an anti-linear ($\SC \, \im\, \SC^{-1}=-\, \im\,$) operation exchanging annihilation and creation operators, $\SC
a^\fd_j\SC^{-1}=u_{\SC,ij}a_j^\dagger$.
\smallskip

\noindent \emph{Invariance of evolution equations --}
 Consider Eq.~\eqref{eq:MarkovianEvolutionGeneral},
for the case where $\hat \rho=\hat \rho(\{a^\fd_i,a_i^\dagger\})$  is the reduced density operator
 describing the state of an open quantum system, and
 $\hat{\mathcal{X}}=\hat{\mathcal{X}}(\{a^\fd_i,a_i^\dagger\})$ the dynamical
 generator governing its out of equilibrium evolution. The transformations
 $\X=\C,\T,\SC$ affecting the fermion operators as $a_i\to \X a_i
\X^{-1}\equiv  (a_i)_\X$ define induced operations $\hat{\mathcal{X}}\to
\hat{\mathcal{X}}_\X$ and $\hat \rho\to\hat \rho_\X$. We
understand an operation $\X$ as a \emph{physical symmetry} if the transformed
evolution equation $\partial_t \hat \rho_\X(t)=\hat{\mathcal{X}}_\X\hat
\rho_\X(t)$ remains  invariant in the sense that 
\begin{equation}
\hat{\mathcal{X}}_\X=\hat{\mathcal{X}}.\nonumber
\end{equation}
In this case, the equation looks the same before and after the
transformation, which implies $\hat \rho_\X(t)=\hat \rho(t)$ for the solutions:  the density operator does not change under the transformation and inherits the symmetry.
\smallskip

\noindent \emph{Anti-unitary symmetry:\! equilibrium vs.\! non-equilibrium --}
Let us now discuss in more specific terms how the anti-unitary transformation $\T$ is represented in the evolution equation. Specifically, we will distinguish between the three different settings mentioned previously, unitary state evolution, equilibrium evolution, and general out-of equilibrium evolution. 

In the first (textbook) case,  $\hat{\mathcal{X}}\stackrel{\mathrm{unitary}}= - \im\,[\hat H,\;. \;]$ generates \emph{unitary time evolution} of the von Neumann equation, $\partial_t \hat
\rho(t)=\hat{\mathcal{X}}\hat \rho(t)$. The combined application of  $\T$ \emph{and} 
an express inversion of physical time, $t\to -t$ describes physical time reversal. The invariance of the equation requires
\begin{equation}
\hat{\mathcal{X}}_\T\stackrel{\mathrm{unitary}}=-\hat{\mathcal{X}},\nonumber
\end{equation}
or $\hat H_\T=\hat H$, where the
$\T$-symmetric Hamiltonian describes equivalent evolution in both time directions ---
the celebrated metaphor of a movie played forward and backward in time. 

The generalization to the case of \emph{thermal equilibrium} is complicated somewhat
by the fact that non-Markovianity becomes essential for fermion systems. Referring for a detailed discussion to section~\ref{sec:beyond_the_Markovian_limit}, the  reason is that the frequency dependent 
Fermi-Dirac distribution coupling dissipation and fluctuation generators via
Eq.~\eqref{eq:FDT} introduces retardation in the time evolution. Within this setting, the previously mentioned  micro-reversibility principle  manifests itself via the  Kubo-Martin-Schwinger (KMS) relation \cite{Kubo1957,Martin1959}
stating invariance of expectation values of any two-time correlators (with time
difference $\Delta t$) under an operation $\Delta t\to -\Delta t  -\im/T$, where $T$
is temperature. This invariance motivates the definition of a generalized time reversal operation acting on functions of time as 
\begin{align}
   \label{eq:EBetaDef}
   \E_\beta f(t)\equiv f(-t-\im\beta).
 \end{align} 
Note that like conventional time reversal  $\E_\beta^2=\mathds{1}$ is an involutory operation. Combined with the Fock space symmetry $\T$, it  defines
the extension of quantum mechanical time reversal to irreversible systems at thermal
equilibrium. The ensuing  \emph{thermal time reversal} operation
\begin{align}
   \label{eq:TBetaDef}
   \T_\beta\equiv\T \circ\E_\beta
 \end{align} 
 is compatible with the presence of a global time arrow in irreversible equilibrium
dynamics.

However, \emph{out of equilibrium dynamics}
excludes changes of the time variable. $\T$ now acts in autonomy as an anti-linear
symmetry realized in Fock space. 
We are thus led to the conclusion that the watershed distinguishing between anti-linear symmetries  with and without time inversion is the boundary between 
unitary or equilibrium dynamics
on the one side, and non-equilibrium dynamics on the other.

What are the consequences of this finding? We first note that both, in and out of
 equilibrium the stationary states, $\hat \rho\equiv \hat \rho(t\to \infty)$ satisfy
 identical symmetries: $\hat \rho =\hat \rho_\X$ is uniquely fixed by the action of
 symmetries in Fock space (we will formulate the ensuing  symmetry classes in more
 concreteness later in the text). This implies a high level of universality for a
 system's state, extending the ten-fold classification of equilibrium phases to the
 full realm of non-equilibrium stationary states. However, both, the dynamical
 generators, and the dynamical processes leading to stationarity satisfy opposite
 symmetry principles. For example, a Hamiltonian contribution, $\hat H$, to a
 non-equilibrium generator must satisfy $\hat H_\T=-\hat H$ to stabilize a
 $\T$-symmetric stationary state, opposite to the equilibrium case. These differences
 are crucially important to the \emph{realization} of stationary states by
 dissipative protocols, a point we will illustrate on the example of a topological
 phase in a chiral symmetry class. An equally important consequence is that in either
 case the language in which the respective symmetry criteria are articulated involves no more than the ten fundamental Fock space symmetries.

Below, we will discuss the above invariance principle for different realizations of
dynamical evolution, including `interacting' $\hat{\mathcal{X}}$s  of quartic order,
and non-Markovian generators required to include the  case of thermal equilibrium. In
either case, the dynamical generators contain three operators, $\hat H,\hat D,\hat P$
describing the effect of unitary state evolution, dissipation and quantum
fluctuations, respectively. These operators are to the non-equilibrium system what
the Hamiltonian is to an isolated quantum system.  As we will discuss in detail, an immediate consequence of this
statement is that  individual (non-Hermitian) operators cannot define a phase, it
takes the more structured information contained in the three operators $\hat H,\hat
D,\hat P$ to do that.

We note that the representation of the anti-linear symmetry $\T$ as a pure Fock space symmetry (out of equilibrium), or in connection with time-reversal as $\T_\beta$ (equilibrium) has consequences for the combined  chiral transformation $\SC = \T\circ \C$ as well. It does not affect, however, the charge conjugation transformation, which is unrelated to time altogether.

\smallskip

\noindent \emph{Gaussian dynamics --} 
 While our approach works for general interacting systems, in the second part of the
paper, we specialize to Gaussian evolutions, the irreversible generalization of free
fermion systems. This setting allows us to present the  general framework in more
concrete terms, and it defines the non-equilibrium counterpart of the free fermion
ground state classification~\cite{Altland97}. The operator $\hat{\mathcal{X}}$ is now
quadratic in $a_i^\fd,a_i^\dagger$, and we may turn to a first quantized
representation in terms of matrices. The deterministic  generator, $\hat K$, is
represented by a non-Hermitian matrix, $K=H-\im D$, with Hermitian contribution $H$,
and semi-positive Hermitian damping matrix, $D$, and the fluctuation generator $\hat
P$ by an anti-Hermitian matrix $P$, see Sec.
\ref{sub:symmetries_in_the_evolution_of_gaussian_states} for the precise relation
between second- and first quantized representation. (Throughout, we use carets to
label all operators in second quantized representation other than
$a_i^\fd,a_i^\dagger$, while first quantized operators come without.) The symmetry
criteria discussed previously now assume the form of matrix symmetries under
transposition or complex conjugation, and give rise to a table with 40 entires,  see
table~\ref{tab2}: 10 symmetry conditions for Hamiltonians in equilibrium, and
$3\times 10$ for the Hamiltonian, dissipation, and fluctuation generators out of
equilibrium, where the number 10 refers to the universal Cartan labels. Coming from a
Hamiltonian perspective, the symmetry relations assume an unfamiliar form, which,
again, has its origin in the absence of time inversion in the present setting.

For a given initial state,  the generator of dynamics determines the system's state
via a   Gaussian density operator. Specifically, we represent the stationary long
time limit  (prior to normalization) as $\rho\equiv \exp(- \Theta )$, where the
Hermitian matrix $ \Theta$ defines the  \emph{effective Hamiltonian}. In equilibrium,
$\Theta = \beta H$ is determined by the Hamiltonian, implying that state and
Hamiltonian transform identically under symmetry operations. Out of equilibrium,
$\Theta = \Theta (H,D,P)$ is a function of the dynamical generator. However, the realization of symmetries on $\Theta$ is determined by the underlying Fock space operations and does not depend on whether the dynamical evolution was in or out of equilibrium (see the last column in table~\ref{tab1}.) For example, the state of a free
fermion topological insulator with `chiral symmetry' is described by a state
anti-commutative with an involutory matrix such as the Pauli matrix, $\sigma_3$, i.e.
$\sigma_3 \Theta\sigma_3=-\Theta$. In equilibrium $\Theta=\beta H$ and  the  Hamiltonian $H$
obeys the same rule. However, if $\Theta$ defines an out of equilibrium distribution defined by a protocol with Hamiltonian participation,   the latter must be commutative, $\sigma_3 H \sigma_3=+H$. The origin of this perhaps counterintuitive result can be traced back to Eq.~\eqref{eq:MarkovianEvolutionGeneral}: the presence or absence of an explicit time inversion in the realization of $\SC=\T \circ \C$ in the equation accounts for the relative sign.

Irrespective of the different realizations of individual symmetries for the generator of Gaussian dynamics, a main
conclusion for  the Gaussian state classification is then  that it leads to the definition
of ten matrix symmetry classes, and that these classes are in one--to--one relation
to the Cartan classes A, AI, AII, \dots , D  defining the `10-fold way', cf. Appendix~\ref{sec:SymmetryClassesTable} for an overview. This correspondence becomes of key importance when we proceed to the classification of
state topologies.

 \smallskip

\noindent \emph{Topology --} 
The objects entering the topological classification are
macroscopic Slater determinants defined by the single particle eigenstates
$|\gamma\rangle$ of Gaussian density operators (or, equivalently, their effective Hamiltonians, $\Theta$), where $\gamma$ indicates the
dependence on parameters such as a lattice momentum or spin-degrees of freedom. We
consider partially filled systems and assume the existence of a subset of states
occupied with high probability, $p_\gamma>\frac{1}{2}$. These  states assume a role analogous to that of quantum ground states, and deviations off
$p_\gamma=1$ describe the residual effects of heating in a system with large
excitation gap. Whether or not the parametric dependence of the states
$|\gamma\rangle$ admits the definition of a topological invariant depends on the
symmetries of $\hat \rho$, and on the  dimensionality of the
parameter space ${\gamma}$. A  corollary of the Cartan classification of
symmetries is that this information can be  lifted from the periodic
table of topological insulators and superconductors \cite{Kitaev09,Ryu10}. For example, a $\Theta$-operator
chiral in the above sense, $\sigma_3 \Theta\sigma_3=-\Theta$, admits the definition of a state
topology in odd but not in even parameter dimension, etc.

Notice that the effective Hamiltonian plays a double role in this context. It defines
topological structures via the parametric dependence of its ground states, and
controls the occupancy of these states via the probabilities
$p_\gamma=f(\epsilon_\gamma)$, where $\Theta|\gamma\rangle=\epsilon_\gamma |\gamma\rangle$
and  $f(\epsilon)=1/(e^\epsilon+1)$. The stable occupancy of the ground state of $\Theta$
requires the presence of a global \emph{purity gap} $\Delta_\mathrm{p}$ \cite{Diehl11,Bardyn2013,Budich2015} defined in Eq.~\eqref{eq:puritygap}, avoiding totally mixed modes with occupation probability $p_\gamma = 1/2$.

As a second condition, we require that the stationary state $\hat \rho=\exp(-\hat\Theta)$ 
is attained with a finite rate, and that excitations out of it relax back in finite
time. 
This permits slow variations of system parameters in time without leaving the instantaneous stationary state.
The dynamical approach towards $\Theta$ is controlled by the deterministic
generator, $K=H-\im D$, and the minimal rate set by the \emph{spectral gap},
$\Delta_\mathrm{s}$, which will be defined in Eq.~\eqref{eq:sgap}. In passing, we
note that the complex eigenvalue spectra  of  dissipation generators are interesting
objects  in their own right (see
Refs.~\cite{bergholtz2019exceptional,ashida2020nonHermitian} for review). Their
singularities in the complex plane, dubbed `exceptional points',  can be classified
in terms of topological principles and  leave signatures in specific dynamical
response functions. However, the  stationary density matrix implies a long time
limit/integral over frequencies effectively averaging over these structures.

We will also address the formation of \emph{edge states} at the boundaries between
bulk topological phases. By definition, an edge is defined by the change in a
topological invariant which in turn requires the closure of the purity gap \cite{Bardyn2013} (unless
the invariant is destroyed by violation of a symmetry condition). Ultimately, one
would like to apply such edges as resources for the realization and manipulation of
topologically protected edge states. In this regard, the closure of the purity gap
appears to be bad news. By definition, a closing purity gap implies  fully mixed
configurations, e.g. Majorana edge qubits forced into an equal probability
configuration of up and down states. However, as we will discuss, a way out of this
dilemma is a simultaneous closure of purity and spectral gap at the edge. We will
see how this happens, for example, in systems with
Lindbladian state evolution where the topological state is the `dark state' of the
dynamics, and how this defines manipulable edge \emph{spaces}.

Finally, we remark that topological classification generally requires more structural input than symmetry classification. For example, for a given fermion Hamiltonian --- classifiable according to the tenfold symmetry scheme --- the objects of topological classification can be zero temperature ground states~\cite{Ludwig_2015} (as in the physics of topological insulators), unitary time evolution operators~\cite{PhysRevB.82.235114,Lindner11} (as relevant for the topology of Floquet systems), or group cohomology structures~\cite{Fidkowski2011} (as relevant for the classification of interacting symmetry protected topological phases). The exploration of the full scope of topological classifications based on the \emph{non-equilibrium} symmetry classification discussed here is a subject transcending the scope of this work.


\smallskip

\noindent\emph{Relation to other work in the field} -- We already mentioned that previous work on symmetry classifications emphasized the non-Hermiticity of non-interacting dissipatively damped dynamics in first quantized matrix representations (encoded in $\hat H$ and $\hat D$, but discarding fluctuations $\hat P$). Building on symmetry classifications of such
non-Hermitian matrices \cite{Leclair02}, it has been reasoned that the Hermitian adjoint $\eta: X \rightarrow X^\dagger$ becomes a symmetry operation in addition to the standard unitary and anti-unitary
symmetries of quantum mechanics \cite{Ueda19,Lee19,ashida2020nonHermitian}. The inclusion of this operation
defined an extended  classification, now containing  $38$  rather than the $10$
classes governing the Hermitian generators of unitary time evolution~\cite{Ueda19,Lee19}. However, causality arguments  led Ref.~\cite{Cooper20} to the conclusion that only ten of these --- defined as combinations of $\eta$, time reversal, $ \mathrm{T}$ and charge
conjugation, $\mathrm{C}$ --- are physical in dissipative evolution. 

 Our analysis of symmetries is different from these works in that  (i) it starts from
 a representation of symmetries in Fock space (as we will see, the passage from
 second quantized operators to matrices is not an innocent one), and (ii) system
 dynamics  is fundamental to the symmetry classification. The latter  requires
 consideration of  all three generators $\hat H,\hat D$ and $\hat P$ on equal
 footing, and on all levels of the description. For example, $\hat P$ contains the
 system's distribution function. If this information is not kept track of, the system
 may appear to be in different classes, depending on whether parts of it are
 integrated out or not. Erasure of this information may thus spoil the unambiguous
 assignment of systems to classes. (The sensitivity of time reversal to the
 realization  of system partitions has also been noted in
 \cite{GinleyCooperNatPh2020}.)

It is important to point out that the symmetry \emph{classes} of  stationary states depend on the symmetries of the above generators but not on those of initial states, unless
multiple stationary states exist. For example, the initial state of a quantum system prepared with definite center of mass momentum breaks $\T$. One may consider fine tuned protocols for which this breaking is preserved including under the evolution by $\T$ invariant generators. However, this situation is not generic. Rather, generic long time evolution will eventually eradicate the memory of the initial state and stabilize a $\T$-symmetric phase. The disclaimer that we are focusing on universal symmetry classes is important inasmuch as non-equilibrium dynamics starting from specific initial states leaves ample room for dynamical or transient  manifestations of symmetries and/or topology not captured by our analysis. As examples, we mention work on the classification of quantum quenches in closed systems governed by topologically non-trivial Hamiltonians \cite{GinleyCooperPRL2018,GinleyCooperPRB2019X,GinleyCooperPRR2019}, and on 
the structure of the complex \emph{spectra} of dissipative generators, and topological signatures caused by the presence of singular `exceptional points' in open system's state evolution \cite{Szameit11,Lee16,Leykam17,kozii2017nonHermitian,Lieu18,Gong18,Yao18,Bergholtz18,Oku19,Rui19,Yoshida19,zirnstein2019bulkboundary,Borgnia2020}, see~\cite{bergholtz2019exceptional} for review. Such structures show in observables probing the approach to stationarity, but not in the stationary phases themselves.

\smallskip

\noindent \emph{Plan of the paper --} The focus in this paper is on general structures.
However, for illustrative purposes
we have included the discussion of one exemplary
case study, namely the physics of an interacting model in class BDI (possessing a
symmetry under both $\T$ and $\C$), reducing to a
variant of a Majorana chain in the Gaussian limit. This example  is deliberately
chosen to illustrate all concepts introduced below in the simplest possible scenario. 

We will start in section~\ref{sec:symmetries} with a discussion of symmetries in Fock space. In section~\ref{sec:symmetries_in_Markovian_dynamics} we discuss the interplay of symmetries and topology in Markovian dynamics. This will introduce the material required to discuss the aforementioned case study in section~\ref{sec:case_study}. Section~\ref{sec:beyond_the_Markovian_limit} goes beyond the Markovian limit and lifts the discussion of symmetries to the framework of the Keldysh path integral. This extension is required, e.g., to include the important limit of thermal equilibrium states. We conclude in section \ref{sec:conclusions}. Technical details are largely relegated to several Appendices.

\section{Symmetries in driven open quantum dynamics} 
\label{sec:symmetries}
In this section, we start out from a precise definition of all symmetry operations $\X$ relevant to our discussion in fermionic Fock space.  This part of the discussion makes no reference to the concrete realization of $\hat{\mathcal{X}}$. In the remaining parts of the paper, we will then explore the manifestations of invariance for the different settings listed in table~\ref{tab1}.


\begin{table}
  \begin{tabular}{|l|l|c|c|}\hline
    & & Lindbladian & Keldysh \cr \hline
    Markovian & linear &$\times$&$\times$\cr
    Markovian & nonlinear &$\times$&$\times$\cr 
    non-Markovian & linear & &$\times$\cr 
    non-Markovian & nonlinear & &$\times$\cr\hline 
  \end{tabular}
  \caption{\label{tab1} Principal types of dissipative dynamics and their theoretical formulations. The inclusion of non-Markovian processes (required, e.g., to describe the relaxation towards a quantum thermal distribution) is beyond the scope of the Lindbladian description but can be addressed within the Keldysh path integral framework.}
\end{table}


\subsection{Symmetry operations in Fock space} 
\label{sub:symmetries}

The symmetry operations relevant to our discussion are defined by $\U,\T,\C,\SC$  (for a review, see \cite{Ludwig_2015,Ryu16}), where  
\begin{alignat}{3}
  \label{eq:UTCSecondQuantized}
  &\U:\qquad&& \U a_i \U^{-1}\equiv u_{ij}\,a_j,\qquad && \U \,\im\,\U^{-1}=+ \im,\cr
  &\T:\qquad && \T a_i \T^{-1}\equiv u_{\T ij}\,a_j, \qquad && \T  \,\im\, \T^{-1}=-\im,\cr
  &\C:\qquad && \C a^\fd_i \C^{-1}\equiv u_{\C ij}\,a^\dagger_j, && \C \,\im\, \C^{-1}=+\im,\cr
&  \SC:\qquad && \SC a^\fd_i \SC^{-1}\equiv u_{\SC ij}\,a^\dagger_j, && \SC  \,\im\, \SC^{-1}=-\im.
\end{alignat}
The
action of the symmetry operations on the creation operators is obtained by taking the adjoint
of the above, e.g., $\T a^\dagger_i \T^{-1}\equiv \bar{u}_{\T
ij}\,a^\dagger_j$, where here and throughout the overbar denotes complex conjugation. 

Eq.~\eqref{eq:UTCSecondQuantized} exhausts the list of unitary ($\U,\C$) and anti-unitary ($\T,\SC$) transformations of the operator algebra exchanging ($\C,\SC$) or not ($\U,\T$) creators and annihilators. The transformations are defined to include an optional purely unitary operation, $u_\X$, where here and in the following $\X = \T,\C,\SC$ unless refined otherwise. Defined as it is, the list contains redundancy. For example, the combination of  $\T$ and $\C$ is a transformation of type $\SC=\T\circ \C$, two different $\T,\T'$ combine to $\U=\T\circ\T'$, etc. However, due to the distinct physical meaning of the transformations, it pays to consider them separately.  

Specifically, $\T$ is in the class of anti-unitary operations required to describe time reversal in unitarily evolving systems. Although the direct meaning of a `time reversing' operation gets lost in irreversible dynamics as anticipated above, anti-unitary transformations continue to play an important role.

The `charge conjugation' transformation, $\C$, exchanges the role of
particle and holes via a \emph{unitary} operation in Fock space, $\C  \,\im\, \C^{-1}=+\im$
\footnote{The terminology `charge conjugation' is a bit of a misnomer, as $\C$ should not be identified with the operation of relativistic quantum physics swapping all positive charges in the universe for negative ones. As with $\T$, the definition of $\C$ above is motivated by its usefulness in the description of irreversible dynamics.}. As mentioned above, the definition of  $\SC=\T\circ\C$ is technically redundant, but considered here for its role in the description of `particle-hole' exchange, and the ensuing `chiral' symmetries.

The group,  $\U$,  includes the familiar unitary symmetries such as  number conservation, $\U=\exp(\im\alpha \hat
N)$, $\hat N=\sum_i a^\dagger_i a^\fd_i$, $\alpha\in \mathfrak{u}(1)$, or spin rotation  $\U_s=\exp(i a^\dagger_\sigma u_{\sigma \sigma'} a^\fd_{\sigma'})$,  $u\in \mathfrak{su}(2)$, ($\sigma,\sigma' = \uparrow, \downarrow$ are spin indices), etc. In the definition of symmetry \emph{classes}, the complementary set of operations $\T,\C,\SC$ must be considered in relation to the unitaries, $\U$. More precisely, these operations
 define meaningful symmetry classes only if they commute with the unitary symmetries of a system, and in the consequence act within the irreducible Fock subspaces defined by them --- sectors of conserved angular momentum, particle number, lattice symmetry, etc. For a more detailed discussion we refer to Refs. \cite{Heinzner05,Ludwig_2015,Ryu16}, and to Appendix \ref{app:unitaryex}, where we discuss this point on a few illustrative examples. 

Finally, we note that all operations 
introduced above are compatible with operator products in that $\X(\hat O\hat O')\X^{-1}=(\hat O\hat
O')_\X=\hat O_\X \hat O'_\X$. This feature will become important when applying transformations to evolution equations such as 
\begin{align}
    \label{eq:SymmetryToEvolutionEquation}
    \partial_t \hat \rho &=\hat{\mathcal{X}}\hat \rho \longrightarrow 
 \X(\partial_t\hat \rho) \X^{-1}  = \X(\hat{\mathcal{X}}\hat \rho)\X^{-1} \cr
  \Longleftrightarrow \partial_t\hat \rho_\X&=\hat{\mathcal{X}}_\X\hat \rho_\X.
  \end{align}  
 \medskip

\noindent \emph{First quantized representation --} To efficiently describe the above transformations in the language of matrices, we define the Nambu operators $A_i\equiv (a^\fd_i,a_i^\dagger)^T$. The operations as defined in
Eq.~\eqref{eq:UTCSecondQuantized}  act on these operators as $A_i
\stackrel{\T}\longrightarrow U_{\T ij}A_j$,  and $A^\fd_i
\stackrel{\X}\longrightarrow U_{\X ij}A_j^\dagger$ (here, $\X=\C,\SC$), where the block diagonal matrix structure
$U_\X=\mathrm{bdiag}(u_\X,\bar{u}_\X)$ is defined in  Nambu block space. For later reference, we note the relations
\begin{align}
  \label{eq:USigmaX}
  U_\X&=\sigma_x \bar U_\X \sigma_x,\cr 
  A^\dagger &= \sigma_x A,
\end{align}
where here and in the following  the Pauli matrices $\sigma_i$ act in Nambu space.
A general bilinear \emph{free fermion operator} $\hat O$  has the Nambu representation
\begin{align}
  \label{eq:FreeFermionOperator}
  \hat O \equiv \tfrac{1}{2}A^{\dagger T}O A\equiv \tfrac{1}{2}A_i^{\dagger} O_{ij}A^\fd_j, 
\end{align}
defined in terms of a  first quantized matrix operator  $O=\{O_{ij}\}$. 
Subjecting  $\hat O$ to the above transformations as
\begin{eqnarray}
  \hat O \stackrel{\X}\longrightarrow \hat O_\X&\equiv&  \tfrac{1}{2}(\X A^{\dagger T}\X^{-1}) (\X O\X^{-1})(\X A\X^{-1})\nonumber\\&\equiv& \tfrac{1}{2} A^\dagger (O_\X) A,
\end{eqnarray}
 (where only $(\T O_{ij}\T^{-1})=\bar{O}_{ij}$ acts non-trivially on matrix elements), we obtain the first quantized version of the symmetry operations (for the relation in the third row, see also App. \ref{app:unitaryex}, Eq. \eqref{Eq:FermiSymmetry}),
\begin{align}
     \label{eq:UTCFirstQuantized}
  \U:\qquad& O \rightarrow O_\U \equiv U O U^\dagger,\cr
  \T:\qquad& O \rightarrow O_\T \equiv U_\T \bar{O}U_\T^\dagger,\cr
  \C:\qquad& O \rightarrow O_\C \equiv -U^T_\C O^T \bar{U}_\C,\cr
  \SC:\qquad& O \rightarrow O_\SC \equiv -U^T_\SC O^\dagger \bar U_\SC.
 \end{align} 
 We say that an operator $\hat O$ has $\X$-symmetry, if $\hat O_\X=\hat O$, or $O_\X=O$ in first quantization. 
 The operations defined by Eq.~\eqref{eq:UTCFirstQuantized} will be the basis for the definition of matrix symmetry classes. 
\smallskip

\noindent \emph{Symmetry tabulation --} Referring to Refs. \cite{Altland97,Heinzner05} for a more comprehensive
discussion, we now briefly identify the symmetry classes defined by application of
$\T,\C$, and $\SC=\C\circ\T$ within the sectors of definite unitary symmetry: two-fold
application of either $\X=\C,\T$ to an $\X$-symmetric operator leads to $O= V_\X O
V_\X^\dagger$, where $V_\T=U_\T
\bar{U}_\T$ and $V_\C=\bar{U}_\C U_\C$.   The commutativity of $V_\X$ with
all operators $O$ in the representation space~\footnote{At this point, the irreducibility of the representation relative to the unitary symmetries of the system is essential. The conclusion $\forall O: [O,V_\X]=0\Rightarrow \X=\pm \mathds{1}$ presumes that $O$ runs through a complete set of matrices in the space where $V_\X$ is defined.} requires $V_\X=\pm \mathds{1}_N$, and
these two options define the symmetry classes $\X=\pm 1$. The absence of
$\X$-symmetry is called $\X=0$. Similarly, the presence/absence of symmetry under $\SC$ is defined as $\SC=0$ or $\SC=1$ \footnote{Two-fold application of $\SC$ defines $V_\SC=(\bar{U}_\SC)^2$. Here, the two options $V_\SC=\pm \mathds{1}$ are unitarily equivalent, $U_\SC \to i U_\SC$, and do not define distinct classes.}. Counting all options, we obtain the list of ten symmetry classes tabulated in Appendix~\ref{sec:SymmetryClassesTable} for the convenience of the reader. 

Already at this stage it is evident that, irrespective of the specific type of the irreversible dynamics (Markovian/non-Markovian and equilibrium/non-equilibrium) we have a maximum number of ten symmetry classes. However, as we will demonstrate below, the conditions under which a given class emerges depend on whether equilibrium or non-equilibrium dynamics is considered.

\section{Symmetries in Markovian dynamics} 
\label{sec:symmetries_in_Markovian_dynamics}

In this section we discuss how the symmetry principles introduced above materialize in the case of Markovian dynamics, i.e. that of systems coupled to a bath with vanishingly short memory. We will start out from a representation of Markovian state evolution in the form of a general Lindblad equation, then specialize to Gaussian state evolution, and finally discuss topological structures.


\subsection{Symmetries in Lindbladian dynamics} 
\label{sub:symmetries_of_lindbladian_dynamics}

To study general Fock space transformations introduced above in the case of Markovian dynamics, we consider a quantum master equation of Lindblad form \cite{Kossakowski1972,Lindblad1976,Gardiner/Zoller,breuer,Alicki2007}
\begin{align}\label{eq:lindblad}
\partial_t  \hat \rho &=    - \im\, [ \hat H  , \hat\rho ]  + \sum_\alpha ( 2 \hat L_{\alpha} \hat\rho  \hat L_{\alpha}^\dagger    -   \{ \hat L_{\alpha}^\dag   \hat L_{\alpha} ,  \hat\rho\}) \equiv \hat  { \mathcal L} \hat \rho.
\end{align}
Here,   $ \hat L_\alpha$ are the `jump operators' coupling the system to a bath, and the sum extends over distinct types of coupling.  Typical realizations include jump operators $\hat L_\alpha\sim a_i$ linear in fermion operators, or number conserving couplings $\hat L_\alpha \sim a_i^\dagger a^\fd_j$.  Number conserving system-bath couplings imply quartic (or even higher order) Lindblad operators,
$ \hat L^\dagger_\alpha \hat L_\alpha=\mathcal{O}(a^4)$, and this is why we need to include nonlinearities even if the focus will ultimately be on Gaussian states.

We now apply the strategy Eq.~\eqref{eq:SymmetryToEvolutionEquation} to the Lindblad
equation, with $\hat \rho\to \hat \rho_\X$ and  $(\im\hat H)\to (\im\hat H)_\X$, $\hat
L_\alpha \to (\hat L_\alpha)_\X$, or $\hat{\mathcal{L}}\to \hat{\mathcal{L}}_\X$ for brevity. The
problem is  $\X$-invariant if 
$\hat{\mathcal{L}}=\hat{\mathcal{L}}_\X$. Under these
conditions $\hat \rho$ and $\hat \rho_\X$ satisfy the same irreversible evolution
equation and are thus identical, $\hat \rho=\hat \rho_\X$. (In cases where the
stationary state depends on the initial state, symmetry of the latter becomes an
additional condition \cite{Albert2014}.) At this level, the analysis includes interacting settings. For
illustration, see the case study of section~\ref{sec:case_study}, where we consider
an interacting one-dimensional chain with  BDI-symmetry, ($\T,\C,\SC)=(+1,+1,1)$.

\subsection{Symmetries of Gaussian states} 
\label{sub:symmetries_in_the_evolution_of_gaussian_states}


While non-linear Lindbladian equations are as complex as their reversible (von
Neumann) cousins, one often has situations where the Hamiltonian is quadratic, i.e.
$\hat H= \tfrac{1}{2}A^{\dagger T}H A$, and Hartree-Fock mean field approximations may
be applied to non-linear operators $\hat L^\dagger_\alpha \hat L^\fd_\alpha$ to
define quadratic approximations (cf. Refs. \cite{Diehl11,Bardyn2013,Tonielli19}).
Under this  condition \cite{Prosen_2008,Prosen_2010,Eisert10}, we have a reduction
\footnote{One has the option to unitarily transform the jump operators to
Lindblad form, $\protect{\hat{L}_{\alpha} = l_{\alpha,i} A_i}$ in the quadratic theory,
such that \unexpanded{$\protect{M_{ij}=\sum_{\alpha}l^\ast_{\alpha,i} l_{\alpha,j}}$}. In this representation,
semi-positivity follows from the properties of the dyadic product.}
\begin{align}
\label{eq:MatrixMDef}
\sum_\alpha  \hat L^\dagger_{\alpha}  \hat  L^\fd_{\alpha} \equiv \hat M \equiv \tfrac{1}{2}A^{\dagger T}M A,   
 \end{align} 
 with a semi-positive and Hermitian matrix $M$, and Hermitian $H$. These conditions reflect the complete positivity of the Lindblad generator \cite{Kossakowski1972,Lindblad1976}, and the dynamical conservation of Hermiticity of $\hat \rho$. The Lindblad equation then assumes the form
\begin{align}
\label{eq:LinbladianM}
\partial_t\hat \rho=-\im\,[\hat H,\rho]- \{\hat M, \hat \rho\}+
2A^T \tfrac{1}{2}\bar M \hat \rho A^\dagger
\end{align}
(similarly to Eq. \eqref{eq:FreeFermionOperator}, $A^T \bar M \hat \rho A^\dagger \equiv A_i \bar M_{ij} \hat \rho A_j^\dagger$). Due to fermion exchange symmetry, the kernel, $O$, defining any bilinear form $\hat O=\tfrac{1}{2}A^{\dagger T}O A$ satisfies $O^T=-\sigma_x O \sigma_x$ (cf. Eq.~\eqref{Eq:FermiSymmetry}). 
We here have assumed without loss of generality that $O$ is traceless.
This feature suggests a decomposition
\begin{align}
  \label{eq:DPDef}
  M&=D - \im P, \,\\ 
   D&\equiv \tfrac{1}{2}(M+\sigma_x M^T \sigma_x), \quad  P\equiv \tfrac{\im}{2}(M-\sigma_x M^T \sigma_x),\nonumber
\end{align}
into contributions symmetric and antisymmetric under Fermi exchange, where the symmetric contribution, $D\geq 0$, inherits the semi-positivity of $M$ \cite{Prosen_2008,Prosen_2010,Eisert10} (this follows from $\langle \psi| M|\psi \rangle\ge 0$ and $\langle \psi \sigma_x| M|\sigma_x \psi \rangle\ge 0$ for any state $|\psi\rangle$). With these definitions we have 
\begin{align}  \label{eq:prop1}
H=-\sigma_x H^T \sigma_x,\quad D=\sigma_x D^T \sigma_x, \quad P=-\sigma_x P^T\sigma_x.
\end{align}
In addition, the  conservation of Hermiticity of the density matrix $\hat\rho$ by the Lindblad generator in each time step implies
\begin{align}\label{eq:prop2}
H^\dag = H,\quad D^\dag = D, \quad P^\dag =- P.
\end{align}

These three building blocks fully describe the dynamical generator, and they all have individual physical meaning: $H$ describes the reversible contribution, $D$ the dissipative damping generator, and $P$ generates fluctuations.  

Eq.~\eqref{Eq:FermiSymmetry} applied to the bilinear forms Eq.~\eqref{eq:MatrixMDef} implies a reduction  $\hat M = -  \im \hat P$ in the  anti-commutator term of the Lindblad equation. However, due to the presence of $\hat \rho$  the full content of $M$ remains in the jump term:
\begin{align}
   \label{eq:LindbladianDPH}
   \partial_t\hat \rho=&- \im (\hat H -\hat P)\hat \rho + \im\hat \rho (\hat H +\hat P) +\cr 
   &+2A^T\tfrac{1}{2}(\bar D +\im \bar P)\hat \rho A^\dagger- c\hat \rho,
 \end{align} 
 with the positive and real constant $c=\tfrac{1}{2}\mathrm{tr}(M)=\tfrac{1}{2}\mathrm{tr}(D)\ge 0$. 

Proceeding as before, we subject Eq.~\eqref{eq:LindbladianDPH} to the operations in Eq.~\eqref{eq:UTCSecondQuantized}, and deduce the symmetry relations the generator compounds $H, D, P$ must obey to obtain form invariance of the evolution equation, and  symmetry $\hat \rho = \hat \rho_\X$ of the density operator. Referring for the details of the  straightforward calculation to Appendix~\ref{sec:appendix_symmetries_of_gaussian_states}, this leads to
the transformation criteria summarized in table~\ref{tab1}.

\begin{table}
\renewcommand{\arraystretch}{1.3}
  \begin{tabular}{|c|c||c|c|c||c||c|}\hline
\multicolumn{2}{|c||}{transformations}&\multicolumn{3}{c||}{non-equilibrium}&\multicolumn{1}{c||}{equilibrium}&\multicolumn{1}{c|}{steady state}\cr\hline
  $\X $& $O_\X$ & $\;\;H\;\;$&$\;\;D\;\;$&$\;\;P\;\;$& $H$ & $\Theta/\Gamma$\cr\hline
  $\T$& $U_\T \bar O U_T^\dagger $&   $-$& $+$ &$- $ &$+$  &$+$\cr
   $\C$& $-U_\C^T O^T \bar U_\C $ & $ +$& $-$  &$+$& $+$ &$+$\cr
  $\SC$& $- U_\SC^T   O^\dagger \bar U_\SC$ & $ -$& $-$ &$ -$ &$+$ &  $+$\cr\hline
  \end{tabular}
  \caption{\label{tab2} Transformation of the (individually Hermitian or anti-Hermitian) matrix operators $O=H,D,P,\Gamma,\Theta$ entering the evolution of the covariance matrix under the fundamental symmetries, $\T,\C,\SC$. The sign factors indicate whether $O=\pm O_\X$ under the transformation. Out of equilibrium, they are chosen to leave the Lindblad equation invariant. In equilibrium (cf. Sec.~\ref{sec:beyond_the_Markovian_limit}, including for the transformation laws of the equilibrium analogs of $D,P$, Eqs. (\ref{eq:HDPEquilibrium}),\ref{eq:HDPEquilibriumS})), the indicated transformation of the Hamiltonian leaves the equilibrium Keldysh action invariant. In either case, the Gaussian stationary states transform as indicated in the last column.  
    }
\end{table}

Notice the similarities and differences to the symmetry classification in unitary
dynamics. Similarly to that case, the ten different combinations are  applied to
operators which are individually Hermitian ($H,D$), or $\im$-times Hermitian ($P$).
However, unlike  with unitary state evolution $\hat \rho \to e^{-\im\hat H t} \hat \rho
e^{\im\hat Ht}$, where just one Hermitian operator determines the symmetry of the
classification object, we here have a situation, where the three operators, $H,D,P$,
are essential to the description of $\hat \rho$, or better to say of the Hermitian
effective Hamiltonian $\Theta$ 
governing the Gaussian states $\hat \rho=\exp(-\tfrac{1}{2}A^{\dagger T} \Theta A)$. In view of the
fact that the focus in the literature (cf. Refs. \cite{Poli15,Zeuner15,Hodaei17,Chen17,Weimann17,Bandreseaar4005,Zhou18,Helbig2020,cerjan19,Xiao20,Cooper20}) often is on the dissipative
damping operator $K\equiv H - \im D$, it is worth pointing out that the symmetry of all
three, $H,D,P$ is essential to the invariance of the state, $\hat \rho$.

Second, we note that due to the lack of the extraneous time inversion $t\rightarrow
-t$ implied in the definition of unitary time reversal symmetry, the symmetries $\T$
and $\T\circ\C=\SC$ are realized differently than in Hamiltonian (or equilibrium)
dynamics. For example, $\T$-symmetry of the density operator requires $H=- U_\T \bar
H U_\T^\dagger$, sign different from the standard time reversal condition of quantum
mechanics. Similarly, the `chiral symmetry', $\SC$, requires commutativity
$H=+\U^T_\SC H \bar U_\SC$, again with opposite sign. In  Sec.
\ref{sub:symmetries_in_systems_with_detailed_balance}, we will compare this symmetry requirement to that in the case  of  unitary evolution.

Finally, notice the absence of a Hermitian adjoint $\eta: O\to O^\dagger$ applied to non-Hermitian operators in
the present framework. To see why, notice that our first quantized $\T$ acts as $\T:
O\to O_{\T}\equiv  U_\T \bar O \U_\T^\dagger 
$ (Eq.~\eqref{eq:UTCFirstQuantized}). By
contrast, Ref. \cite{Cooper20} suggested $\T': O\to O_{\T'}\equiv U_{\T'}  O^T \U_{\T'}^\dagger$. The two
transformations differ by a relative application of $\eta$, $\T'= \T\eta$ (up to
unitaries). The $\T'$ transformation was motivated by a criterion of causality --  namely to avoid a sign change in the imaginary contribution to the damping generator $K=H-\im D$ -- and by requiring a smooth connection of the irreversible dynamics to purely unitary  evolution.

In our approach, the causality/sign issue does not arise since the transformation $t\to -t$ is avoided. This is seen by noting that $K$ couples to the dynamics  (symbolically) as $\exp(-\im K t)=\exp((-\im H-D)t)$. The absence of the  $t\to -t$ operation in
our understanding of symmetries in irreversible non-equilibrium dynamics makes $\eta$ never appear in any symmetry
operation.

\subsection{Topology of Gaussian states} 
\label{sub:topology_of_gaussian_states}

Turning to topology, the \emph{effective Hamiltonian} governing $\hat
\rho=\exp(-\tfrac{1}{2}A^{\dagger T} \Theta A)$ becomes center stage \cite{Bardyn2013,Rivas2013,Viyuela1D,Viyuela2D,Arovas2D,Budich2015,Linzner2016,Bardyn2018,ZhangGong2018,Bandyopadhyay2020} (see \cite{Coser2019} for an alternative approach to mixed state topology). The  topological states defined
by $\hat \rho$ are the `ground states' of $\Theta $, i.e.  Slater determinants formed from
the set of all eigenstates, $\{\psi_\gamma^-\}$ with negative eigenvalues,
$\epsilon_\gamma$. Before discussing the topological structure of these states, let
us investigate how the symmetries introduced above extend to those of $\Theta $.

Rather than probing $\Theta $ directly, we here consider the stationary \emph{covariance matrix}, $\Gamma=\lim_{t\rightarrow\infty}\tilde \Gamma(t)$ \cite{Prosen_2008,Prosen_2010,Eisert10}, which carries the same information but is more directly accessible. The covariance matrix is a $2N\times 2N$ Hermitian matrix defined as 
\begin{align}
  \label{eq:CovarianceMatrix}
 \tilde \Gamma_{ab}(t)\equiv \mathrm{tr}(\hat \rho
  [ A_a,A_b^\dagger]), 
\end{align}
where $a$ is a composite index comprising the Hilbert space index, $i$, and the two-component Nambu index.
Inspection of $\Gamma$ in the eigenbasis shows that
 \begin{eqnarray}
 \Gamma= \tanh(\Theta /2),
 \end{eqnarray} 
 i.e. the covariance matrix and the effective Hamiltonian carry identical information, and in particular share the same ground state. Substitution of Eq.~\eqref{eq:CovarianceMatrix} into Eq.~\eqref{eq:lindblad} shows that $\Gamma$ is obtained as the stationary solution of the equation
  \begin{align}
    \label{eq:GammaEqOfMotion}
    \partial_t \tilde\Gamma(t)&=-\im K \tilde\Gamma(t)+\im  \tilde\Gamma(t) K^\dagger - 2 \im P, 
  \end{align}
  $K\equiv H - \im D$.  The long time limit of the solution is obtained as
\begin{align}
  \label{eq:GammaFrequencyRepresentation}
  \Gamma&= - 2\im\int_{-\infty}^\infty \frac{d\omega}{2\pi}\frac{1}{\omega-K}P \frac{1}{\omega-K^\dagger}\cr
  &= - 2\im \int_0^\infty dt e^{-\im Kt}Pe^{+\im K^\dagger t}. 
\end{align}
 Eq.~\eqref{eq:GammaFrequencyRepresentation} reveals much of the physics
  of the covariance matrix, and the dynamical processes stabilizing it. Specifically,
  the equation shows that the stationary state is obtained by retarded/advanced
  propagation of the fluctuation matrix $P$, as described  by the retarded and
  advanced Green functions  $(\omega-K)^{-1}$ and $(\omega-K^\dagger)^{-1}$. The
  non-negative matrix $D$ contained in $K=H-\im D$ defines the relaxation rate at which the
  stationary state is attained. We require a finite minimal rate, as defined by the
  spectral gap, 
  \begin{eqnarray}\label{eq:sgap}
  \Delta_\mathrm{s}= - \mathrm{min}\bigl(\text{Im}\, [\mathrm{eigenval}(K)]\bigr) >0.
  \end{eqnarray}
  %
This requirement permits slow (adiabatic) changes of system parameters in time such that the system stays at all times in the
instantaneous steady state.
Note that only the imaginary part of the eigenvalues, i.e. the lifetime of the  slowest decay rate in the system, enters the damping gap; thus the Hamiltonian alone may be zero, or have zero modes without corrupting an open damping gap.

Finally, it is straightforward to check either by inspection of the representation Eq.~\eqref{eq:GammaFrequencyRepresentation}, or of the evolution equation Eq.~\eqref{eq:CovarianceMatrix}, that the symmetries  of $K, H,P$ listed in table~\ref{tab2} extrapolate to symmetries of $\Gamma$ and $\Theta$ as indicated in the last column.  

On this basis, we now discuss the topological structures defined by $\Theta$. This operator plays a role analogous to the gapped single particle Hamiltonians describing topological insulators or superconductors. 
 The dissipative analog of the single particle gap in these systems is the purity gap \cite{Bardyn2013,Budich2015}, which can be defined via the eigenvalues of $\Gamma = \tanh (\Theta/2)$ as
\begin{eqnarray}\label{eq:puritygap}
\Delta_\mathrm{p}=\min\bigl(| \mathrm{eigenval}(\Gamma)|\bigr) >0.
\end{eqnarray}
%
 The purity gap sets a lower bound for the  negative part of $\Theta $'s spectrum relative
 to $0$. In terms of $\Theta $, a vanishing purity gap is similar to a `metal', where a chemical potential intersecting
 a band invalidates the definition of a topological ground state. The occupation probability of an $\Theta $-eigenstate $\psi_\gamma$  is
 given by the Fermi function $p_\gamma = f(\epsilon_\gamma)=1/(1+e^{\epsilon_\gamma})$. This
 occupation balance demonstrates that negative eigenvalues of finite minimal  modulus
 are required to stabilize a fully occupied ground state in the thermodynamic limit
 containing a macroscopically large number of states. In contrast, a state with vanishing purity gap features at least one mode with $\epsilon_{\gamma_*} = 0$, representing a totally mixed fermion state with occupation probability $p_{\gamma_*} =1/2$. 

For systems with finite spectral and purity gap, the ground state $\{\psi_\gamma\}$ of $\Theta $ defined by the Slater determinant of all negative eigenvalue $\Theta $ eigenstates becomes the subject of topological classification. Since the symmetries of $\Theta $ are realized identically to those describing Hermitian Hamiltonians, the classification of dissipative
 topological phases becomes equivalent to that described by the periodic table of
 topological insulators (cf. table~\ref{tab:PeriodicTable}). The same goes for physical information obtained via topological
 principles from the periodic table. For example, a two-dimensional dissipative Chern
 insulator in class $\mathrm{A}$ ($(\T,\C,\SC)=(0,0,0)$) \cite{Budich14,Goldstein2019,Tonielli19,Shavit2020} supports circulating chiral
 edge states, or a dissipative quantum wire in class $\mathrm{BDI}$
 ($(\T,\C,\SC)=(+1,+1,1)$), has Majorana end states \cite{Diehl11}, etc.

\subsection{Edge state formation} 
\label{sub:edge_state_formation_}

While the classification of bulk topological phases requires the presence of a purity gap, at the boundary this gap closes. In fact, one may take the closing of the purity gap as a definition of the phase boundary. This interpretation follows from the identification of the covariance matrix with a band `Hamiltonian', and its ground state as the carrier of a topological index. Changes in the ground state require the closure of a gap, presently the purity gap in the spectrum of $\Gamma$. Edge states are the low lying eigenstates of $\Gamma$ spatially confined to that boundary region.

While the existence of edge states is a robust feature granted by topology, the
 accessibility and manipulability of these states --- a feature required by, e.g.,
 quantum information applications --- is another matter.
 In fact, we here run into the seemingly paradoxical situation, where the closure of
 the gap around zero eigenvalue in $\Gamma$, i.e. the definition of the edge appears
 to contradict the accessibility of these states. To see how, consider the density
 matrix $\hat \rho =\exp(-\tfrac{1}{2}A^{\dagger T}\Theta  A)$ projected to the edge state subspace.
 Assuming $\Theta $ to be diagonalized, $\Theta =\mathrm{diag}(\{\epsilon_\gamma\})$ individual
 states $|\gamma\rangle \equiv a_\gamma^\dagger |0\rangle$ are occupied with
 probability $f(\epsilon_\gamma)$. For $\epsilon_\gamma\simeq 0$, the Fermi function
 approaches the value $1/2$, which means that the state of the edge is given by a
 mixed state defined by equally occupied and empty edge states. For example, a
 Majorana wire whose two-Majorana edge space would be in an equal weight mixture of
 its two states, and hence useless for `qubit' applications.

However, there is a loophole in the argument. It presumes that the \emph{spectral} gap remains open at the boundary. To see why this matters, consider the solution of the differential equation \eqref{eq:GammaEqOfMotion} governing the evolution of the covariance matrix,
\begin{align}\label{eq:timeev}
  \tilde \Gamma(t)=\Gamma_0(t) - 2\im \int_0^t ds \,P(t-s),
\end{align}
with $O(t)\equiv e^{-\im K t}O e^{\im K^\dag t}$, and initial state $\Gamma_0$. If the spectral gap, defined as in
 Eq.~\eqref{eq:sgap}, is finite, the stationary state $\Gamma$ of
 Eq.~\eqref{eq:GammaFrequencyRepresentation} is approached exponentially fast. However,
 now consider the different situation, where the approach of the boundary goes along with a simultaneous
 vanishing of all three generators of the dynamics, $H,D,P$ within the subspace in which the purity gap closes. In this case,  the solution of
 the evolution equation projected to that space will retain information on the initial state, and we have an edge capable of
 storing information. The condition of simultaneous vanishing of all generators is not quite as stringent as it may seem. For example, it is realized in dissipative systems possessing a \emph{dark state} or multiple of these forming a \emph{dark space}. A dark state is a state $\rho=|\Psi\rangle \langle \Psi|$ stationary under  the full generator of
 the Lindbladian dynamics Eq.~\eqref{eq:LinbladianM}. This condition requires the simultaneous vanishing of all partial generators within the dark space. For an example and the discussion of the ensuing edge space, see section \ref{sub:edge_states}.

\section{Case study} 
\label{sec:case_study}

\begin{figure}
\centering
\includegraphics[width=8.5cm]{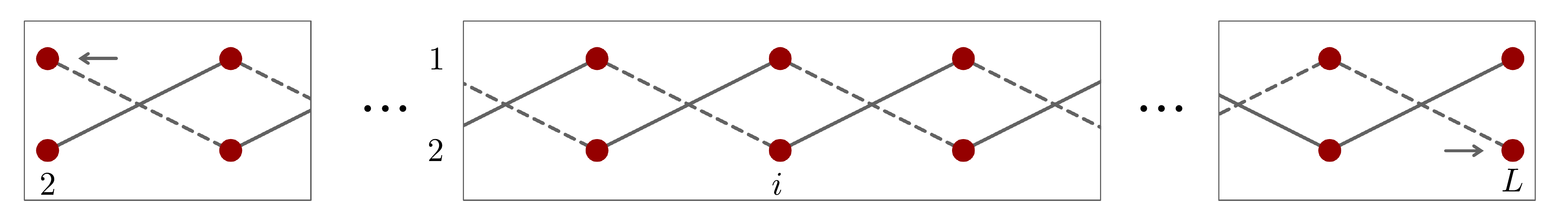}
\caption{\label{fig:SSHModel} 
Class BDI one-dimensional lattice model containing two competing hybridization processes indicated via solid and dashed lines. The model has a topological stationary state, whose winding number changes in a topological phase transition at the degeneracy point of equal hybridization strength.
 }
\end{figure}

In this section, we consider a one-dimensional lattice model in the non-equilibrium symmetry class
BDI $(\T,\C,\SC)=(+1,+1,1)$ for $H, D$, and $P$ to illustrate the general concepts introduced above, and
connect to various other concepts currently discussed in the literature.  The model
is defined by a  chain of $L$ sites $i$ containing two orbitals $1,2$
indicated by red dots in Fig.~\ref{fig:SSHModel}. For the moment, we assume periodic boundary conditions, a system with edges will be considered below. The  irreversible dynamics is governed by number-conserving jump operators, so  that the Lindbladian is quartic in
fermion operators. We consider a half filled system in which a Hartree-Fock style
linearization of the evolution around a macroscopically filled state is possible, following the construction principles of \cite{Diehl11,Tonielli19}. The
ensuing mean field dynamics is controlled by a parameter $\vartheta$ such that for
$\vartheta=0$ the stationary state is a product state of decoupled equal weight superposition (spin-$x$) states
defined along the solid links  in the figure. In the opposite extreme,
$\vartheta=\pi/2$, it is a spin-$x$ state defined via the dashed lines. For generic
values, the eigenstates of the effective Hamiltonian $\Theta$ are spatially extended, and the
full ground state is characterized by a winding number which changes in a topological
phase transition at $\vartheta=\pi/4$. In the following, we discuss both purely
dissipative protocols stabilizing this state, and generalizations including an  added
Hamiltonian.

\subsection{Interacting model} 
\label{sub:interacting_model}

 Let $a_{1,i}$ and $a_{2,i}$ be   annihilation operators for orbitals $1$ and $2$ at site $i$, and define the two-component operator $a_i=(a_{1,i},a_{2,i})^T$.  Now consider the set of transformed two-component operators defined as 
%
%
%
\begin{eqnarray}
\hat l_i &=& \frac{1}{\sqrt{2}}\left(
\begin{array}{c}
a_{1,i+1}+a_{2,i}\\
-a_{1,i+1}+a_{2,i}
\end{array}\right),\nonumber\\
\hat r_i &=& \frac{1}{\sqrt{2}}\left(
\begin{array}{c}
a_{1,i}+a_{2,i+1}\label{eq:SSHOperatorLDef}\\
-a_{1,i}+a_{2,i+1}
\end{array}\right).
\end{eqnarray}
acting on neighboring lattice sites. $\hat l_i$ and $\hat r_i$ are related to the original operators by unitary transformations
$ \hat  l_i=V_La_i$ and $\hat  r_i=V_Ra_i$, with $V_L =e^{i \frac{\pi}{4}\Sigma_y}(E^{11}\hat\tau  +E^{22})$ and 
 $V_R =e^{i \frac{\pi}{4}\Sigma_y}(E^{11} +E^{22}\hat \tau)$, such that $\{\hat l_{a,i}\}$ and likewise $\{\hat r_{a,i}\}$ ($a=1,2$ denoting the orbital index) generate
  a fermion algebra as well. 
 Here $(E^{ij})_{kl}=\delta^i_{\,k}\delta^j_{\,l}$ are matrices in orbital space, 
    $\Sigma_i$ are Pauli matrices in the same orbital space,
  and $\hat \tau$ is a lattice translation operator $\hat \tau a_i = a_{i+1}$. 
  %
  %
  
  To get some intuition for  the operators $\hat l$, consider the $N$ particle configuration defined by occupation of all
  $\hat l_{2}$ states: $|L\rangle \equiv \prod_i \hat  l_{2,i}^\dagger |0\rangle$. Since we have
  $2N$ sites in total, this is a half filled state. With
  $\hat l_{2,i}^\dagger=\frac{1}{\sqrt{2}}(-a_{1,i+1}^\dagger+a_{2,i}^\dagger)$ we observe
  that it is a product state defined by an equal weight superposition (or spin
  $x$-state) defined along the solid bonds in the figure. Conversely,  $|R\rangle \equiv \prod_i \hat r_{2,i}^\dagger |0\rangle$ is a product state of spin-$x$ hybridizations along the dashed bonds. Second the states $|L\rangle,|R\rangle$ afford an interpretation of free fermion ground states of distinct topological order.  To see how, consider the operators $\hat H_L\equiv \sum_i \hat l_i^{\dagger ^T}\Sigma_z \hat l^\fd_i$. The state $|L\rangle$ is the gapped ground state of this Hamiltonian at half filling. On the other hand, substitution of the definition \eqref{eq:SSHOperatorLDef} shows that $\hat H_L = \sum_q a_q^{\dagger T}\left(\begin{smallmatrix}
      &w^\dag \cr w &
    \end{smallmatrix}    \right)a_q^\fd$, with $w = e^{\im q}$ in momentum space.
  This comparison identifies $|L\rangle$ as the ground state of a gapped chiral
 parent Hamiltonian, where the winding of the phase   $z= \exp(\im q)$  as a function of
 $q\in [0,2\pi)$ defines a topological invariant  $W=\frac{1}{2\pi \im}\int dq z^{-1}
 \partial_q z= 1$. Likewise, the state $|R\rangle$ is the ground state of a
 Hamiltonian $\hat H_R$ with winding $W=-1$.

In the following, we construct dissipative  protocols driving into the states $|L\rangle, |R\rangle$, or more generally a competition between them. Depending on the balance at which these reference states enter we expect different topological phases, with a transition between them. Specifically, it will be instructive to 
 consider two different Lindblad protocols ($c = \cos \vartheta, s = \sin \vartheta$), 
\begin{align}
  \label{eq:SSHLIncCoh}
\hat{\mathcal{L}}_{\mathrm{inc}}&\equiv c\hat{\mathcal{L}}_{ l} + s \hat{ \mathcal{L}}_r,
  \cr
\hat{ \mathcal{L}}_{\mathrm{coh}}&\equiv \hat{\mathcal{L}}_{c l+s r},
\end{align}
where $\hat{\mathcal{L}}_x$ is the Lindblad operator defined relative to the fermion operator choice $\hat x$, (i.e. $\hat l, \hat r$, or $c\hat l+ s\hat r$ respectively). 
%
%
The $\hat{\mathcal{L}}_{x}$ are bilinears in jump operators $\hat L$ as in Eq.~\eqref{eq:lindblad}. Suppressing the site index we define the latter as 
\begin{align}
  \label{eq:SSHJumpOperators}
  \hat L_{\ell1} = \sqrt{\kappa} a_\ell^\dagger \, \hat x_1,\qquad \hat L_{\ell2} = \sqrt{\kappa} a_\ell \, \hat x_2^\dagger,
\end{align}
where $\ell=1,2$, $\hat x=\hat l,\hat r $, or $\hat x=c \hat l+s \hat r$ for the second protocol, and $\kappa$ setting the coupling strength. Note that these operators are number conserving and act locally in real space. Specifically, 
 the jump operator $\hat L_{a1}$ depletes the upper $x$-band and $\hat L_{a2}$ fills the lower, so that the joint action of the two fills the ground state associated to the $\hat x$-operators. The jump operators $\hat L_{an}$, $a,n=1,2$, define a Lindblad equation Eq. \eqref{eq:lindblad} with $\hat H=0$ and the sum extending over the four configurations and lattice sites, $\alpha\equiv (a,n,i)$. 

The `incoherent' operator  $\hat{ \mathcal{L}}_{\mathrm{inc}}$ puts the two operators cooling into $|L\rangle$ and $|R\rangle$, respectively, into competition, while the operator $\hat{\mathcal{L}}_{\mathrm{coh}}$ cools into the ground or 'dark' state defined by a coherent superposition of $\hat l$ and $\hat r$. Either way, we expect a phase transition at $\vartheta = \pi/4$, where $c=s=\frac{1}{\sqrt{2}}$ and the left and the right operator algebra couple at equal strength. 
%
%

\subsection{Symmetries} 
\label{ssub:symmetries}

Turning to symmetries, we first note that the  Lindblad operators in the $\hat{\mathcal{L}}_x$ model are $\T=1$ symmetric with auxiliary unitary $U_\T=1$ in the sense of our earlier discussion. The reason is that $\bar V_{L,R}=V_{L,R}$ are real, and hence $\T \hat L_{an} \T^{-1}= \hat L_{an}$ for real $\kappa,c,s$.  

Second, we have a  $\SC=1$ symmetry under the chiral symmetry operation
\begin{align*}
  \SC a^\fd_i \SC^{-1}\equiv \Sigma_z a_i^\dagger,
\end{align*}
with auxiliary matrix $U_\SC=\Sigma_z$.
Noting that the matrices $V_{L,R}$  
in Eq.~\eqref{eq:SSHOperatorLDef} satisfy the symmetry 
\begin{align}
  \label{eq:SSHVSym}
  \Sigma_z V \Sigma_z=(-\im\, \Sigma_y)V,
\end{align}
it is straightforward to verify that 
\begin{align}
  \SC \hat L_{a n} \SC^{-1}= (-1)^{a+1} \hat L_{a\bar n},
\end{align}
where $\bar n = 2, 1$ for $n = 1,2$. 
Since the $\hat L$'s appear pairwise and are summed over, the sign factor drops out, and $\mathcal{L}_x$ has a chiral symmetry.

Finally, the composition $\T\circ \SC$ induces a $\C=+1$ symmetry under the \emph{unitary} map $\C a^\fd_i \C^{-1}\equiv \Sigma_z a_i^\dagger$, with $U_\C=\Sigma_z$. We thus have a model with BDI symmetry $(\T,\C,\SC)=(+1,+1,1)$ on the second quantized level.

\subsection{Linearized model} 
\label{sub:linearized_model}


 $\hat{\mathcal{L}}_x$ is quartic in fermion operators and hence defines a nonlinear
problem. However, this nonlinearity can be Hartree-Fock decoupled, thanks to the
macroscopically large number of particles in the problem. To see how, consider the
bilinears (no sum convention here) $\hat L_{a1}^{\dagger}\hat L_{a1}=\kappa
(\hat x^\dagger E^{1a}a^T)(a^{\dagger T}E^{a1}\hat x)= \kappa \hat x_1^{\dagger} a_a\,a_a^{\dagger
}\hat x_1 \to  \kappa \hat  x_1^{\dagger} \langle a_a a_a^{\dagger }\rangle
\hat x_1=\frac{\kappa}{2}\hat x_1^{\dagger}\hat x_1$ at half filling $\langle a_{ai} a_{ai}^{\dagger
}\rangle=1/2$, used in the final step. Likewise, $\hat L_{a2}^{\dagger}\hat L_{a2}\to
\frac{\kappa}{2}\hat x^\fd_2\hat  x_2^{\dagger}$.  With $\hat x=Va$, and $V$ the linear
transformation defining the site operators $\hat x$ via $a$ (e.g. $V=V_R$ in
Eq.~\eqref{eq:SSHOperatorLDef} for $\hat x=\hat r$), we thus find
\begin{align}
  \sum_\alpha \hat L^\dagger_\alpha \hat L^\fd_\alpha\equiv \hat M \to \frac{\kappa}{2}A^{\dagger T} \left(\begin{matrix}
   V^\dagger E^{11} V^\fd & \cr & V^T E^{22}\bar V 
  \end{matrix} \right)A,
\end{align}
where $A=(a,a^\dagger)^T$  as before. Comparison with Eq.~\eqref{eq:DPDef} shows that the principal building blocks of the dynamical generator are given by
\begin{align}
  \label{eq:SSHPDDef}
  D=\kappa \left(\begin{matrix}
    V^\dagger V &\cr & V^T \bar V 
  \end{matrix}  \right), \quad P= \im\kappa\left(\begin{matrix}
    V^\dagger \Sigma_z V &\cr &- V^T \Sigma_z \bar V  
  \end{matrix}  \right),
\end{align}
where we rescaled $\kappa\to 4\kappa$ for notational simplicity. Since there are no terms
coupling  the particle and hole sector in Nambu space, we focus on the
particle blocks throughout, and redefine $D\to \kappa V^\dagger V$, and $P\to \im\kappa
V^\dagger \Sigma_z V$ for simplicity. Using Eq.~\eqref{eq:SSHVSym}, we note that in
this first quantized representation, the chiral symmetry acts as $D\to
D_\SC=-\Sigma_z D \Sigma_z=-D$, and $P\to P_\SC=+ \Sigma_z P \Sigma_z= - P$. This is
in accordance with table~\ref{tab2} and implies that $D$ and $P$ are diagonal and
off-diagonal in orbital space, respectively. Time reversal  manifests itself in the
reality of the matrices $D$ and $\im P$.

In passing, we note that this reality can be used to represent the system in the \emph{Majorana representation} of the BDI chain \cite{Kitaev01}. For example, with $\eta=\frac{1}{\sqrt{2}}(a^\dagger +a)$, $\nu =\frac{1}{\im\sqrt{2}}(a^\dagger - a) $, it is straightforward to verify that $a^{\dagger T} D a=- \im \eta^T D \nu $. However, the representation change will not play an essential role throughout.

We now have everything together to explore the stationary states stabilized by this dynamics. As a warm-up exercise, consider the case $\hat x=\hat l$. The corresponding matrices $V_L$ are unitary, and a quick calculation using Eq.~\eqref{eq:SSHOperatorLDef} yields the momentum representation
\begin{align}
\label{eq:SSHPL}
   D=\kappa\mathds{1},\qquad P=\im\kappa (E^{12} e^{-\im q}+ \mathrm{h. c.}),
 \end{align}
 with matrix structure in orbital space. Substitution of these expressions into Eq.~\eqref{eq:GammaFrequencyRepresentation} defines the momentum representation of the covariance matrix 
\begin{align}
  \label{eq:CovarianceLeftWinding}
  \Gamma_L\equiv \bar z_L E^{12}+ z_L E^{21},  
  \qquad  z_L=e^{\im q}.
\end{align}
We thus conclude that the system cools at uniform rate $(D=\mathrm{const.}\mathds{1})$ into the $|L\rangle$ state with its winding number $W=1$. The covariance matrix (and, equivalently, the effective Hamiltonian, $\Theta$) are chiral in the sense of table ~\ref{tab2}, such that we have a dissipatively realized class BDI
system with stable topological phase.

Next consider the incoherent driving protocol. Since the Lindblad generators are linear in the matrices $P$ and $D$,  we now have a linear superposition, $D=\kappa (c+s)\mathds{1}$, and $P=\im\kappa (
E^{12}(c t^\dagger + s t)+\mathrm{h. c.})
$. Substitution of these expressions into Eq.~\eqref{eq:GammaFrequencyRepresentation} gives
\begin{align}
  \Gamma_{\mathrm{inc }}\equiv \bar z_\mathrm{inc} E^{12}+  z_\mathrm{inc} E^{21}, 
   \quad z_\mathrm{inc}=\frac{c e^{\im q} + s e^{-\im q}}{c+ s}.
\end{align}
The straightforward computation of the winding number shows that $W=\mathrm{sgn}(c-\frac{1}{\sqrt{2}})$. At the phase transition, $c=s=\frac{1}{\sqrt{2}}$, the purity gap closes at two linearly dispersive points $q=\pm \pi/2$. (The presence of two Dirac points in a two-band model corresponds to the change of the winding number by $2$ at the phase transition.) However,the damping gap, i.e. the smallest eigenvalue of $D$, remains open everywhere.

Now, compare this to the drive $\hat{\mathcal{L}}_x$ defined by the linear superposition $\hat x=c  \hat l + s  \hat r$. Up to a normalization factor, the algebra of $\hat x$-operators satisfies canonical commutation relations, $\{ \hat x^\fd_{a,q},\hat x^\dagger_{a',q'}\} =
\delta_{aa'}\delta_{qq'}(1+2sc\cos q)$, implying that we cool into a pure state. Proceeding as above, we define a matrix $V=c  V_L +
s  V_R$, which is `almost' unitary, $V^\dagger V= (1+ sc
(t+t^\dagger))\mathds{1}$, reflecting the scaled commutation relations. Using Eq.~\eqref{eq:SSHPDDef}, we find $D=\kappa(1+ 2sc
\cos q)$, and $P=\im\kappa (E^{12}(c  e^{-\im q}+s )^2 e^{\im q}+ \mathrm{h. c.})$, which gives
\begin{align}
  \Gamma_{\mathrm{coh}}&\equiv \bar z_\mathrm{coh} E^{12} +  z_\mathrm{coh} E^{21}, \cr 
  & \quad z_\mathrm{coh}=  \frac{c^2 e^{\im q}+s^2 e^{-\im q}+2sc }{1+ 2sc
\cos q}. 
\end{align}
Again, we have a topological phase transition at $c=s=\frac{1}{\sqrt{2}}$. However, its critical signatures are markedly different from those of the incoherent case. 
The covariance matrix now has unit-norm eigenvalues, $|z_{\text{coh},q}| = 1$, reflecting the purity of the stationary state. Interpreted as a topological band insulator Hamiltonian, it thus resembles a Hamiltonian with `flattened' non-dispersive spectrum. The corresponding purity gap equals unity, except at the phase transition point, where the absence of a well defined limit in $z_{\text{coh},q}$ as $q \to\pm \pi$ reflects the singularity required to have a unit norm curve $z_{\text{coh},q}$ change its winding number around the origin. However, the damping gap, generally open in the incoherent case, now does close at criticality.

\subsection{Edge states} 
\label{sub:edge_states}

Both, in the coherent and the incoherent protocol, an edge is defined by the closure
of the purity gap. In the incoherent protocol, the spectral gap remains open, and in
the light of the discussion of section~\ref{sub:edge_state_formation_} this implies a
non-manipulable edge in the sense that the edge covariance matrix will 
approach $\Gamma=0$ exponentially fast, reflecting a fully mixed state. However, in the coherent
protocol the situation is more interesting. Here, too, a (now singular) vanishing of the purity gap defines the edge. However,  this
singularity goes along with a vanishing of both $D$ and $P$. 

The situation can be described both, in a continuum representation via a smooth
variation of the parameters $(c,s)$ through the critical point, or directly on the
lattice. We here choose the latter option, and identify a simultaneous null-space of
the operators $D$ and $\im P$ in real space, for a system cut open as indicated in
Fig.~\ref{fig:SSHModel}. To this end, we first identify  a vector $|\Psi_0\rangle$
annihilated by $V$. With $V=cV_L+sV_R$ and $V_{L,R}$ given by
Eq.~\eqref{eq:SSHOperatorLDef}, $V|\Psi_0\rangle=0$ 
is equivalent to the two conditions 
$c \psi^1_{i+1}+s \psi^1_i=0$, and $s \psi^2_{i+1} + c \psi^2_i=0$. For $c>s$ this
recursion relation has an exponentially decaying solution $\psi^1_i = \psi^1_1
\left(-\frac{s}{c}\right)^{i-1}$, $\psi^2_i=0$ centered on the upper orbital of the
left edge, and a partner solution $\psi^2_i=\psi^2_L\left(-\frac{s}{c}\right)^{L-i}$
at the lower orbital of the right edge, see the arrows in the figure. (For $c<s$, the
role of upper and lower orbital are reversed.) The decoupling of these states from
the system is obvious in the case $c=1$, where the eigenstates are
local hybridizations indicated by the solid bonds, excluding  the states $(n=1,i=1)$ and $(n=2,i=L)$.
The definition \eqref{eq:SSHPDDef} implies that $|\Psi_0\rangle$ is a simultaneous
eigenstate of $P$ and $D$.

For unitary systems whose Hamiltonian has the same topology as $\Gamma_\mathrm{coh}$,
these states are the Majorana edge states of the BDI chain (the limit $c=1$
corresponding to its `sweet spot', where the decoupling of an edge Majorana is
manifest in the lattice representation). Presently, the  same topological setting
manifests in a dark space, spanned by $|\Psi_0\rangle \langle \Psi_0|$,
$|\Psi_1\rangle \langle \Psi_1|$, where the many body states $|\Psi_{0,1}\rangle$
differ in the occupation of the complex fermion corresponding to the two Majoranas
centered at the left and right edge, respectively. The decoupling of $|\Psi_0\rangle$ 
from the dynamics implies that arbitrary mixed states $\hat \rho=
\sum_{i=0,1}p_i|\Psi_i\rangle \langle \Psi_i|$, $p_0+p_1=1$ are solutions of the
dynamical equations and we have a freely manipulable edge space.


\subsection{Adding a reversible contribution} 
\label{sub:adding_a_reversible_contribution}


 While our so far discussion focused on purely dissipative dynamics, it is straightforward
to include a Hamiltonian contribution $H_q = h \hat h_q \cdot \vec \Sigma$, where
the unit vector $\hat h_q$   describes the action of the Hamiltonian in orbital
space, and $h$ is its strength  (here assumed uniform for
simplicity.)
 For definiteness, we study the influence of this Hamiltonian on a purely
left winding dissipative background, $\hat x= \hat l$. Parameterizing the dissipative
generators~\eqref{eq:SSHPL} associated to the Lindblad operators in Eq.
\eqref{eq:SSHOperatorLDef} in the same way, $D_q = \kappa \mathbb 1$, and $P_q =
\im \kappa \hat p_q \cdot \vec \Sigma$, where  $\hat p_q =  (\cos q,\sin q,0)^T$,  it is
straightforward to verify that the stationary covariance matrix assumes the form
 \begin{eqnarray}\label{eq:ourfirstGamma}
 \Gamma_q  =& \,\, \hat p_q \cdot \vec \Sigma  - \frac{h}{\kappa^2+h^2} \left(\kappa + h \hat h_q \times\right) (\hat p_q\times \hat h_q)\cdot \vec \Sigma .  
 \end{eqnarray}
 This solution demonstrates the influence
of the Hamiltonian on the stationary state. In general,  only Hamiltonians obeying the symmetry
condition of table~\ref{tab2}, $H=+\Sigma_z H\Sigma_z$, or $\hat h_x=\hat h_y=0$ define a $\Gamma$-matrix chiral in the sense $\Sigma_z \Gamma \Sigma_z=-\Gamma$. In
other words, it takes a Hamiltonian \emph{diagonal} in the orbital space to preserve
the off-diagonality of $\Gamma$. (One may trace the origin of this perhaps unexpected
finding back to the fact that the anti-unitary irreversible chiral symmetry operation
$\SC=\T\circ\C$ does not involve a sign change of physical time.) 
If in contrast $H$ would obey the equilibrium relation $H=-\Sigma_z H\Sigma_z$, chiral symmetry of the steady state would in general be lost.
We also note  that a symmetry preserving $H$ does not commute with $P$ (the
Pauli matrix structure again) and hence degrades the purity of the state: for non-vanishing coupling $h$, the eigenvalues of $\Gamma$ no longer have unit modulus. While the purity gap gets affected, the spectral gap, $\kappa$, remains untouched, as long as $[D,H]=0$, which the setup above assumes.
  
Finally, there is one interesting exception to the general rule above: in cases, where the dynamics stabilizes a \emph{dark state}, the symmetry condition on $H$ gets lifted. A dark state is a zero eigenstate of the full Lindblad generator. Translated to the language of the matrices, $H,P,D$, this requires commutativity of all three of them, and $P^2=D^2$. While the latter has been a feature built into our model from the outset, the commutativity $[H,P]=0$ implies $\hat h_q\times \hat p_q=0$, and hence the vanishing of the second term in Eq.~\eqref{eq:ourfirstGamma}. In this way, the Hamiltonian decouples, and we are left with the pure and chiral configuration described by the first term, i.e. the projector onto the dark state. 

\subsection{Non-Hermitian SSH model and exceptional points} 
\label{sub:nonHermitian_ssh_model_and_exceptional_points}


Against this background, let us address a few generalizations of  Hermitian one-dimensional systems with chiral symmetry \cite{SSH}, which have been discussed in the recent literature. Specifically, the  non-Hermitian SSH model \cite{Lee16,Lieu18,zirnstein2019bulkboundary}  is defined by the matrix 
\begin{align}\label{eq:nhSSH}
H_\mathrm{nh} &=    \left(\begin{array}{cc}
0  & z_+t^\dag\\
  z_- t & 0 
\end{array}
\right)\\\nonumber 
&  =\left(\begin{array}{cc}
0  & z e^{-\im q}\\
  \bar z e^{\im q} & 0 
\end{array}
\right) + \left(\begin{array}{cc}
0  & \Delta z e^{-\im q}\\
  -\overline{\Delta z} e^{\im q} & 0 
\end{array}\right) \\\nonumber 
 &\equiv H_1 +\im \, H_2 ,
\end{align}
where $z_{\pm}=z\pm \Delta z$ are 
arbitrary complex numbers. Can this model feature
as a generator of dissipative dynamics, be associated to a symmetry class, or define a fermionic topological state?

Turning to the first part of the question, we need to understand  how
$H_{\mathrm{nh}}$ relates to the general triptych $H,D,P$ of unitary, dissipation and
fluctuation generators, respectively. Since the imaginary parts of the eigenvalues
$\lambda_{\pm}= \pm\sqrt{z_+ z_-}$ have indefinite sign, the direct identification $H_\mathrm{nh}\stackrel{?}=H - \im D$  does not define a legitimate dissipation generator. As a cure, one could add an overall unit matrix $-\im \kappa'\mathds{1}$, with $\kappa' \geq |\mathrm{Im}\lambda_{\pm}|$ to make it sign definite, such case is considered below. Here, we discuss the alternative interpretation $H_\mathrm{nh}= H_1 +\im \, H_2=H+P$ as the sum of a Hermitian part
$H_1\equiv H$ and a fluctuation generator $H_2\equiv -\im P$. Choosing $\Delta z\equiv
\im \kappa$ for definiteness, a quick calculation shows that the fluctuation generator
$\im H_2=P$ becomes identical to the matrix $P$  of the winding protocol
Eq.~\eqref{eq:SSHPL}. However, the  definition of a consistent Lindbladian dynamics
requires the balancing presence of a dissipation generator $D$ of matching
strength such as $D=\kappa\mathds{1}$. We are led to the conclusion that $H_\mathrm{nh}$ by itself does not define a consistent dynamical protocol, while the generator defined by the 
matrices $H=H_1$, $D=\kappa \mathds{1}$ and $P=\im H_2$ does. Referring back to the discussion of the previous section, this generator stabilizes the covariance matrix \eqref{eq:ourfirstGamma}. However, this
matrix, and the corresponding effective Hamiltonian, $\Theta$, do not have chiral symmetry since $H_1$ does not fulfill the non-equilibrium chirality condition of table
\ref{tab2}.
Hence $\Theta$ does not possess
topological ground states. Although $H_\mathrm{nh}$ `looks chiral', it does not define a fermionic topological state of Lindbladian dynamics. 
\smallskip

\emph{Role of exceptional points} -- Exceptional points are terminal points of branch cut singularities forming in the complex spectra of non-Hermitian linear operators \cite{Heiss_2012}. At these points eigenvalues merge, and topological numbers counting the multivaluedness of the corresponding eigenvalue can be determined. A prototypical setup of this sort is defined by a  non-Hermitian matrix $N=N_1+\im N_2$ with Pauli matrix structure and  Hermitian and anti-Hermitian contributions, $N_1 = n_{1,\mu} \Sigma_\mu, N_2 = n_{2,\mu} \Sigma_\mu$, where $\mu=0,1,2,3$, $\Sigma_0=\mathds{1}$, and real coefficients $n_{1,\mu}, n_{2,\mu}$. The semi-positivity constraint reads $n_{2,0}\geq |\vec n_2| \geq 0$. Its eigenvalues
\begin{eqnarray}
\lambda_\pm = n_{1,0} + \im n_{2,0} \pm \sqrt{(\vec n_1 + \im \vec n_2)^2}, 
\end{eqnarray}
merge at the terminal point of the branch cut defined by the vanishing of the argument of the square root.
 The classification of exceptional points according to the symmetries of their parent operators, $N$, has attracted a lot of attention in the recent literature \cite{Szameit11,Lee16,Leykam17,kozii2017nonHermitian,Lieu18,Gong18,Yao18,Bergholtz18,Oku19,Rui19,Yoshida19,zirnstein2019bulkboundary}. 

Do exceptional points affect the phases forming as stationary limits of driving
protocols? To answer this question, we need to take a look at  non-Hermitian linear
combinations of the three constituent operators ($H,D,P$). Specifically, we have seen
that $K=H-\im D$ governs the approach towards the stationary limit.
 It is straightforward to realize
exceptional points by adaption of the chiral model dynamics studied above. 
As an example, consider the matrices $V_L$ in \eqref{eq:SSHOperatorLDef} modified as $V_L \to V_L (E^{11}+r E^{22})$ with real $r$. Inspection of Eq.~\eqref{eq:SSHPDDef} shows that this changes the matrix $D=d_\mu \Sigma_\mu$ as $d_0= \frac{\kappa}{2}(1+r^2), \vec{d}=\frac{\kappa}{2}(1-r^2)\hat{e}_z$. Adding to this the Hamiltonian contribution $H=H_1$ of the SSH model Eq.~\eqref{eq:nhSSH}, we obtain a model with dissipation generator $K=H-\im D$, whose eigenvalues are given by
\begin{align*}
  \lambda_{\pm}=-\frac{\im\kappa}{2}(1+r^2)\pm \left(|z|^2-\frac{\kappa^2(1-r^2)}{4}\right)^{1/2}.
\end{align*}
Both eigenvalues have negative imaginary part, a necessary condition to define a valid dissipation generator. For $|z|= (1-r^2)^{1/2}\kappa/2$ an exceptional point with degenerate eigenvalues is realized.

Does this non-analyticity affect the nature of the system's stationary phases? To
understand what is happening, consider the general formula for the covariance matrix
Eq.~\eqref{eq:GammaFrequencyRepresentation}, represented as a frequency integral over
the retarded and advanced propagators $G^R \equiv (\omega-K)^{-1}$ and $G^A \equiv
(\omega-K^\dagger)^{-1}$. With $\mathrm{Im}\lambda_\pm<0$ in the lower complex plane,
$G^{R/A}$ is analytic in the upper/lower half of the complex plane, with simple poles
at $\omega=\lambda_\pm$ and $\omega=\bar\lambda_\pm$, respectively. At first sight,
it looks like the residue of the pole integration might depend on the values
$\omega=\lambda_\pm$ with their non-analytic dependence on system parameters.
However, actually doing the integrals one finds that the resulting expression for the
covariance matrix is a \emph{rational function of these parameters}.

In other
words, the presence of exceptional points has no effect on the stationary long-time
phase. 
While we have no proof demonstrating this feature in the most general terms,
it is straightforward to verify  for translationally invariant dynamical generators
with two internal bands, and we suspect it to be of general nature. 
Non-analyticities in the complex eigenvalue spectra of dissipation generators may affect the transient dynamical stages on the way towards the stationary limit or the dynamical response of the stationary state 
\cite{Poli15,Weimann17,Zeuner15,Zhou18,Hodaei17,Chen17,cerjan19,Xiao20,Helbig2020,Bandreseaar4005}. The physical significance of exceptional points hosted in the non-Hermitian matrix $K$ for the  dynamical evolution is clearly read off Eq.~\eqref{eq:timeev}, in the perhaps clearest way for a system that approaches an infinite temperature state ($P=0$). However, they do not appear to affect the ensuing stationary phases and their topological classification in themselves. 

 \section{Beyond the Markovian limit} 
 \label{sec:beyond_the_Markovian_limit}

 Our discussion so far focused on the Markovian case of a memoryless environment.
 This excludes important settings, notably those stabilizing quantum thermal
 distributions of fermion systems. The extension to non-Markovian situations is achieved via the Keldysh
 path integral formalism \cite{FEYNMAN1963,Keldysh1964}. In this section, we will
  discuss how the apparatus of symmetries manifests itself in the path integral, and then apply it to situations outside the Markovian limit. The main subject of this section is the extension of the time reversal transformation
 of quantum mechanics, Fock space $\T$ combined with time
 inversion $t \to -t$, to irreversible equilibrium dynamics. The ensuing
 thermal time reversal transformation, Eq.~\eqref{eq:TBetaDef} is essential to the identification of the symmetries of  $\hat H, \hat D,\hat P$ in equilibrium settings. Naturally, the symmetries of $\hat H$ coincide with those of the standard zero temperature setting familiar from the literature. However, they differ  from those found for
 non-equilibrium dynamics, table~\ref{tab2}.

 \subsection{Symmetries in the Keldysh path integral} 
 \label{sub:symmetries_in_the_keldysh_path_integral}

Within the fermion Keldysh framework, physical observables are computed as expectation values of a coherent sate path integral (see Refs. \cite{Kamenev2011,Altland/Simons,Sieberer2016} for review). Specifically, the covariance matrix  now assumes the form 
 \begin{align}
 \label{eq:KeldyshFunctional}
  \Gamma_{ab}(t)=\int \, e^{\tfrac{\im}{2}\int \frac{d\omega}{2\pi}\, \eta^T_{-\omega}( \omega \tau_1 -\M_\omega)\nu_\omega}\,\nu^c_{a}(t) \eta^c_{b}(t)\cr
\end{align}
where 
the kernel 
\begin{align}
  \label{eq:KeldyshKernelMDef}
  \mathcal{M}_\omega=\left(\begin{matrix}
   0 &K_\omega^\dagger\cr K_\omega& - 2P_\omega 
  \end{matrix}  \right)
\end{align}
may have  time, $\mathcal{M}=\mathcal{M}(t-t')$, or equivalently frequency dependence. The functional integration $\int \equiv \int D(\eta,\nu)$ is over two Grassmann variables
$\eta=(\eta^c,\eta^q)^T$, $\nu=(\nu^c,\nu^q)^T$  where  each
component  $\eta^{c,q}=(\eta_1^{c,q},\eta_2^{c,q})^T$ is subject to a Nambu doubling, and carries
a Hilbert space index. 
 The
Pauli matrices $\tau_i$ act in $c/q$ space. $\eta^{c/q}$ are often referred to as 'classical' (c) and 'quantum' (q) components of the fields.
We note that complex conjugation has no meaning for these integration
variables. However, with the Fourier convention ($(d\omega)=\frac{d\omega}{2\pi}$)
\begin{align*}
  \xi(t)\equiv \int (d\omega) \, e^{-i\omega t}\xi_\omega,\qquad \xi=\eta,\nu,
\end{align*}
we have $\overline{\xi}_\omega=\xi_{-\omega}$, as for real variables. 

The Keldysh path integral contains the information previously expressed in terms of the covariance matrix. To see how, we use the rules of Gaussian integration to obtain
\begin{align}
  \label{eq:KeldyshPropagator}
  &\langle \nu_{a,\omega}^c\eta_{b,\omega'}^c\rangle=\delta(\omega-\omega') \left(\frac{1}{\omega \tau_1-\mathcal{M}}\right)_{ab}^{cc}=\\ 
  &\qquad=\delta(\omega-\omega')\left(\frac{1}{\omega-K_\omega} (- 2 \im P_\omega) \frac{1}{\omega-K_\omega^\dagger}\right)_{ab}. \nonumber
\end{align}
For frequency independent $P_\omega = P$, $K_\omega =K$, this reduces
to  the covariance matrix Eq.~\eqref{eq:GammaFrequencyRepresentation} of the Markovian framework. However, the advantage of the Keldysh path integral is that it allows us to go beyond that, and include processes with memory. Referring for a more detailed discussion to section~\ref{sub:symmetries_in_systems_with_detailed_balance}, the most important representative of this category is the thermal Fermi-Dirac distribution. In this case, the damping and fluctuation kernels are related by the fluctuation dissipation relation Eq.~\eqref{eq:FDT}, which implies that the process remains non-Markovian ($P$ cannot be approximated by a constant) at all frequencies, $\omega$.    

The practical identification of symmetries within the path integral formalism  differs somewhat from our previous strategy. There, we had asked what symmetry transformations leave equations of motion invariant. Presently, it is more natural to ask what transformations leave the path integral action unchanged. Since equations of motion are  implied by the path integral (even though they may assume an effectively intractable form in non-Markovian situations), the two strategies are equivalent.
Within the present approach, we will use various freedoms in manipulating the integral, notably the option to exchange the order of variables. We here illustrate these features on a simple consistency check, namely the path integral verification of the Hermiticity of $\Gamma$. This introduces the manipulations required in the subsequent discussion of anti-unitary symmetries. Expressed in path integral 
language, 
\begin{align*}
   &\Gamma^\dagger_{ab}(t)=\overline{\Gamma_{ba}(t)}=\overline{\langle \nu^c_{b}(t) \eta^c_{a}(t)\rangle}\cr 
    &=\int \, e^{-\tfrac{\im}{2}\int (d\omega)\, \eta^T_{\omega}(\omega \tau_1 -\overline{\M_{\omega}})\nu_{-\omega}} \nu^c_{b}(t)\eta^c_{a}(t)\cr 
    &=-\int \, e^{\tfrac{\im}{2}\int (d\omega)\,  \nu^T_{-\omega}(\omega \tau_1 -(\M_\omega)^\dagger)\eta_{\omega}} \eta^c_{a}(t)\nu^c_{b}(t)\cr 
    &=\int \, e^{\tfrac{\im}{2}\int (d\omega)\,  \nu^T_{-\omega}(\omega \tau_1 -\M_{\omega}) \eta_{\omega}} \nu^c_{b}(t)\eta^c_{a}(t)=\Gamma_{ab}(t).
\end{align*}
In the second line, we took the transpose in the action (catching a minus sign due to the anticommutativity of the Grassmann fields), and in the third  used
$  \M_\omega^\dagger=-\tau_3 \M_{\omega} \tau_3$ and a variable transform $\eta \to \tau_3 \eta$,  $\nu\to -\tau_3 \nu$ , 
 to arrive at an expression identical to the original integral, except differently named integration variables.

Similar manipulations applied to the covariance matrix transformed as indicated in the last column of table \ref{tab2} (for details, cf. Appendix \ref{sec:symmetries_in_keldysh_representation}) lead to the condition ($U_\X$ act as unit matrices in Keldysh space)
\begin{alignat}{3}
\label{eq:SymmetriesKeldyshKernel}
  &\T:\qquad && \mathcal{M}_\omega=-\mathcal{M}_{\omega,\T}&&=-U_\T\overline{\M_{-\omega}} U_\T^\dagger,\cr
  &\C:\qquad && \mathcal{M}_\omega=+\mathcal{M}_{\omega,\C}&&= - U^T_\C\M^T_{-\omega} \bar U_\C,\cr
   &\SC:\qquad && \mathcal{M}_\omega=-\mathcal{M}_{\omega,\SC}&&=  U^T_\SC(\M_\omega)^\dagger \bar U_\SC.
\end{alignat}
Substitution of Eq.~\eqref{eq:KeldyshKernelMDef} into these relations yields conditions for the blocks, $P$, $K=H-\im D$ identical to those listed in  table~\ref{tab2}. However, recall that for a matrix kernel with time dependence, complex conjugation $\overline{\M_\omega}=\overline{\M}_{-\omega}$ implies a change in the frequency variable. 

\subsection{Symmetries in systems with detailed balance} 
\label{sub:symmetries_in_systems_with_detailed_balance}

 Above we saw that the stabilization of a $\T$-symmetric effective
Hamiltonian, $\Theta=\Theta_\T$ from a dynamical process containing a Hamiltonian
contribution, $H$, requires the Hamiltonian to transform as $H\to - H_\T$. But how
can this be reconciled with the familiar case of thermal equilibrium, 
 where a $\T$-symmetric effective Hamiltonian $\Theta=\beta H$ forms via thermalization
(likewise an irreversible process) of a system with $\T$-symmetric $H$? The
resolution to this seeming paradox lies in a point mentioned in the introduction,
namely that the $\T$ operation appropriate to the description of Markovian
irreversible dynamics leaves the physical time parameter, $t$, untouched. By
contrast, physical time reversal in unitarily evolving systems is described by an
operation $\T_0 \equiv \T\circ \E_0$, where $\E_0$ is the operation $\E_0: t\to -t$ in a
`theory space' containing $t$ as an external parameter. Systems at thermal
equilibrium are not time reversal symmetric in the strict sense of unitary evolution.
However, they obey a principle of `micro-reversibility', or detailed balance, which
makes the reflection of time a meaningful operation.

To see how this comes about, consider the correlation function $\mathrm{tr}(\hat \rho\hat A_t\hat B_{t'})$
with Heisenberg evolved operators $\hat A_t=e^{\im\hat H t}\hat A e^{-\im\hat H t}$, and thermal, normalized $\hat \rho =\mathcal{Z}^{-1}e^{-\beta \hat H}$  with partition sum $\mathcal{Z}$. We work in second quantized representation, as indicated by the carets. As a straightforward consequence of the cyclic invariance of the trace, one obtains the Kubo-Martin-Schwinger (KMS) relation \cite{Kubo1957,Martin1959},
\begin{align}
\label{eq:KMSRelation}
  \mathrm{tr}(\hat \rho\hat A_t\hat B_{t'})=\mathrm{tr}(\hat \rho\hat B_{t'}\hat A_{t+\im\beta}),
\end{align}
i.e. an operator reordering relative to $\hat \rho$ at the expense of a shift of the time parameter into the complex plane. For Heisenberg evolved operators, this equation is based on the specific form of $\hat \rho$, and hence is a signature of the thermal state. Time reversal plays no role, up to now. 

Now use the relation 
\begin{align}
  \label{eq:TDaggerTrace}
  \mathrm{tr}(\hat {\Theta})=\mathrm{tr}(\hat {\Theta}^\dagger_\T)
\end{align}
 to fuse the KMS relation with an anti-linear transformation in Fock space. (In a first quantized representation, the auxiliary relation follows from $\mathrm{tr}({\Theta}_\T^\dagger)=\mathrm{tr}(U_\T \bar {\Theta}^\dagger U_\T^{-1})=\mathrm{tr}({\Theta}^T)=\mathrm{tr}({\Theta})$. For a verification in second quantized representation, see Appendix~\ref{eq:AppYTrace}). A straightforward combination of Eqs.~\eqref{eq:KMSRelation} and~\eqref{eq:TDaggerTrace} then leads to 
\begin{align}
  \label{eq:KMSCombinedWithT}
  \mathrm{tr}(\hat \rho\hat A_t\hat B_{t'})=\mathrm{tr}(\hat \rho(\hat A_\T)_{-t-\im\beta}  (\hat B_\T)_{-t'},
\end{align}
where we assumed $\T$-invariance $\hat H_\T=\hat H$ of the Hamiltonian.

This equation states the invariance of two-time correlation functions under a simultaneous application of the Fock space anti-linear transformation $\T$ and the  transformation $\E_\beta$ (Eq.~\eqref{eq:EBetaDef}) acting on functions of time
$E_\beta f(t) = f(-t-i\beta)$.
 Provided that $\hat A_\T=\hat A$ and $\hat B_\T=\hat B$ are $\T$-invariant, the combined operation $\T_\beta=\T\circ \E_\beta$ of Eq.~\eqref{eq:TBetaDef}  
 defines a symmetry of the correlation function.  Notice that the operation $\T_\beta $
 is the product of two transformations $\T$ and $\E_\beta$ acting in different
 spaces, namely Fock space and functions defined over Fock space, respectively. For
 $\beta\to 0$ we obtain the standard operation of quantum mechanical time reversal,
 i.e. anti-linear $\T$ followed by an inversion of time. Here, the `infinite
 temperature limit' simply means that in this case, the symmetry makes a statement
 for unrestricted traces of operators. 
 
 The above symmetry of correlation functions under thermal time reversal $\T_\beta$
 suggests that this operation might define a  symmetry of the theory in general. To
 verify this expectation, we consider the Keldysh functional~\cite{Altland10,Sieberer15,Crossley17,Glorioso2017,Haehl2017b,Aron18} and analyze how its
 action transforms under $\T_\beta$.  We
 do not assume a thermal distribution just yet, but will obtain it as part of the
 criteria required for invariance.  In Keldysh language, the inversion of time
 amounts to (a) an exchange of the forward ($+$) and the backward ($-$) time contour,
 (b) a contour dependent sign due to the reverse ordering of fermion/Grassmann fields
 along the contour, summarized jointly as 
$\nu_\pm \to \pm \nu_{\mp}$, where $\nu_{\pm}=(\nu^c \pm \nu^q)/\sqrt{2}$ are the fields on the contours $\pm$, and (c) an inversion of the time parameter $\nu(t)\to \nu(-t)$, or $\nu_\omega \to \nu_{-\omega}$ (which effectively changes the sign of all terms odd in frequency parameters such as $\int (d\omega) \eta^T_{-\omega}\omega \nu_\omega$). In the $\nu=(\nu^c,\nu^q)^T$ representation via `classical' and `quantum' fields, the combined effect of these operations assumes the form  $\xi_{\omega}\to
(\im\tau_y)\xi_{-\omega}$,  $\xi=\nu,\eta$. This needs to be
supplemented by (d) a shift by  $\im\beta$ into the complex time domain. In the 
frequency representation,  the shift operation on the individual contours assumes the form of  a multiplicative factor $\exp(\pm \omega \beta/2)$. Turning to the $(\nu^c,\nu^q)$ form, we obtain the full representation of $\E_\beta$, as 
\begin{align}
  \label{eq:EbetaKeldyshFields}
  \E_\beta:\xi_\omega&\to \mathcal{Y}_\omega \xi_{-\omega},\qquad  
  \mathcal{Y}_{\omega}=(\im\tau_y) \exp\left(\frac{\beta\omega}{2}\tau_x\right),
\end{align}
for both $\xi=\nu,\eta$.
Having established the representation of $\E_\beta$ on Keldysh fields, we now need to figure out what effect $\T_\beta = \T\circ \E_\beta $  has  on a Keldysh action with kernel $\mathcal{M}$ as in Eq.~\eqref{eq:KeldyshKernelMDef}, with generally frequency dependent $K_\omega=H_\omega-\im D_\omega$, and $P_\omega$. 

Referring to Appendix~\ref{sec:proof_of_eq_xx} for details, we find that the theory is invariant under thermal time reversal $ \T_\beta$ provided that (i)
\begin{align}
\label{eq:HDPEquilibrium}
  H_{\omega,\T}=H_\omega,\qquad D_{\omega,\T}=D_\omega, \qquad P_{\omega,\T}=-P_\omega,
\end{align}
where the notation emphasizes, that the symmetry not only tolerates, but actually requires frequency dependence of the operators $P$ and $D$. The latter is constrained by condition (ii)
\begin{align}
  \label{eq:FDT}
    P_\omega &=  \im \tanh\left(\frac{\beta\omega}{2}\right)D_\omega, 
\end{align}
which is an implementation of the fluctuation-dissipation relation.
This identity conditions fluctuations ($P$) and dissipation ($D$) to each other, via a frequency dependent factor which one may consider the definition of the global equilibrium temperature.

Notice how the conditions for $\T$ differ from those listed for the non-equilibrium case in table \ref{tab2},
with sign changes owed to the active transformation of the time parameter. By
contrast, the transformation $\C$ does not relate to the time parameter in either
case and remains unchanged. However, the combined transformation $\SC
=\T\circ \C$ inherits the changes from $\T$,
\begin{align}
\label{eq:HDPEquilibriumS}
  H_\SC=H,\qquad D_{\omega,\SC}= - D_\omega, \qquad P_{\omega,\SC}= -P_\omega.
\end{align}
The unmodified transformation laws under $\C$, together with the modified ones in
Eqs. (\ref{eq:HDPEquilibrium},\ref{eq:HDPEquilibriumS}), yield the equilibrium column
of table \ref{tab2}, generalizing the standard symmetry classification
\cite{Altland97,Heinzner05} to finite temperature settings. Reflecting the fact that for Gaussian systems in equilibrium $\Theta=\beta H$, these conditions are identical to those in the third column $(\Theta/\Gamma)$ of table \ref{tab2}. 
While our analysis was performed for a Gaussian setting, it is natural to expect that the same symmetries characterize the self energies $\Sigma(D,P)$ forming in an interacting system relaxing into an equilibrium configuration at arbitrary temperature $T$. In this case, the symmetries are inherited from that of the microscopic parent theory under Eq.~\eqref{eq:TBetaDef}, where $\T$ acts on Fock space operators and $\beta$ is set by the temperature of a background bath determining the system's temperature. 
(The role of this bath can be played by the system itself, in which case the temperature is determined by the energy of the initial state from which the thermalizing evolution departs.)

We finally note that,  as in the complementary out of equilibrium
case, two gaps stabilize a topological phase: a spectral gap $|\mathrm{eigenval}(H)|
>0$, and a purity gap in $\Gamma_{\text{eq}} = \tanh \beta H/2$, realized for
temperatures smaller infinity, $\beta >0$, as long as the spectral gap is open \cite{Bardyn2018}.

\subsection{Scope of the equilibrium symmetry conditions} 
\label{sub:scope_of_the_equilibrium_symmetry_classes}

Our discussion above showed that their combined symmetry under Fock space operations,
$\X$, and generalized time reversal, $\E_\beta$, defines the symmetries of
micro-reversible systems different from that of out of equilibrium systems. This
makes one wonder just how general the scope of the micro-reversible framework is. Can
it be extended beyond the category of equilibrium systems?

We first note that a \emph{symmetry class} in the sense of our present discussion is
defined by a set of operators $(H,D,P)$ sharing a certain set of conditions under
application of $\X=\T,\C,\SC$. (The realization of the symmetry in the dynamical
evolution may or may not include an additional transformation of time under
$\E_\beta$.) Individual deformations of these operators do not leave the symmetry
class, provided they do not violate the symmetries of the constituent operators. This
is an important disclaimer. For example, a lattice Hamiltonian invariant under the
combined application of lattice inversion and time reversal does not define a class,
because the addition of static disorder would violate the symmetry.

In this reading,  the irreversible relaxation into an equilibrium configuration does
not define a class, as it requires the `fine tuning' $D \propto P$,
Eq.~\eqref{eq:FDT}. Physically, a configuration $D \propto P$ builds up at long times
when a system acts as its own thermalizing bath; before reaching that stage,
traces of the initial state --- possibly with different symmetry properties --- may remain visible. In this sense, $D\propto P$ defines an attractive
`surface'  in the `space' $H,D,P$, provided the conditions for thermalization are
met.

To make this point more concrete, we consider the long time limit of the covariance matrix, obtained from Eq.~\eqref{eq:KeldyshPropagator} as
\begin{align}
  \label{eq:CoviKeldyshExplicit}
 \Gamma&=-2\im \int_{-\infty}^\infty \frac{d\omega}{2\pi}\frac{1}{\omega-H+\im D }  P_\omega \frac{1}{\omega-H-\im D}\cr 
 &= -2\im \int_{-\infty}^\infty \frac{d\omega}{2\pi} G^+_\omega  P_\omega G^{-}_\omega ,
\end{align}
where in the second line we emphasize the dynamical interpretation of the covariance matrix by defining the retarded and advanced propagators
\begin{align}
  G^\pm_{\omega} \equiv (\omega - H \pm \im D)^{-1}.
\end{align}
For simplicity, we neglect optional frequency dependences in $H$ and $D$, but in view of Eq.~\eqref{eq:FDT} not in $P_\omega$. 
  
On this basis, we ask under what circumstances does the covariance matrix  have symmetry
under $\T$. \emph{Out of equilibrium}, the answer is given by the first three columns
of table \ref{tab2}. To see this in explicit terms, compute $
\Gamma_\T=U_\T^\fd
\overline \Gamma U_\T^\dagger$ as
\begin{align*}
  \Gamma_\T= +2\im \int_{-\infty}^\infty \frac{d\omega}{2\pi} G^-_{\omega,\T}  P_{\omega,\T} G_{\omega,\T}^{+},
\end{align*}
where $G_{\T,\omega}^\pm \equiv (\omega - H_\T + \im D_\T)^{-1}$. 
Provided the
symmetries hold as stated, a change of variables $\omega\to -\omega$ brings us back
to the original expression, $\Gamma_\T=\Gamma$. We repeat that this symmetry requires oddity $H_\T=-H$ of the system Hamilton operator.

\emph{In equilibrium}, we have the `fine tuning' Eq.~\eqref{eq:FDT} which implies a different option to establish $\T$: In this case, the numerator may be written as $-2\im P_\omega =2\tanh(\frac{\beta \omega}{2})D=\im\tanh(\frac{\beta \omega}{2}) [(G^-_\omega)^{-1}-(G^+_\omega)^{-1}]$, and the covariance matrix assumes the form
\begin{align}
   \Gamma \stackrel{\mathrm{eq.}}= \im\int \frac{d\omega}{2\pi}\tanh \left(\frac{\beta\omega}{2}\right)(G^-_\omega - G^+_{\omega}),
 \end{align} 
 i.e. an integral over the spectral function of the system (in general broadened by
 the coupling to the bath establishing equilibrium) over the Fermi-Dirac distribution
 function. In this case, $\Gamma=\Gamma_\T$ holds if $H=+H_\T$, as required by the
 equilibrium column of table Eq.~\eqref{tab2}.

However, notice that this symmetry crucially relies on the proportionality $D\propto P_\omega$, Eq.~\eqref{eq:FDT}. It gets broken by even mild departures away from equilibrium realized, e.g. by coupling to two baths kept at different temperatures, $T_1$ and $T_2$. In this case, we have $D=D_1+D_2$ which in general is no longer proportional to $P_1 +P_2 = \im \tanh(\frac{\beta_1 \omega}{2})D_1+\im\tanh(\frac{\beta_2 \omega}{2})D_2$. Of course, the out of equilibrium symmetry relation still holds, provided $H,D,P$ satisfy the required criteria.

This also provides us with the opportunity to point out that both in and out of equilibrium, we never encounter the Hermitian adjoint, $\eta$, as a natural part of symmetry operations. For example, this operation would act on our Green functions as $G^+_\omega =(\omega-H+\im D)\stackrel{\eta}\longrightarrow G^-_\omega=(\omega - H -\im D)$. Considering the time representation of these propagators, 
\begin{align}
  G^\pm(t)=\int \frac{d\omega}{2\pi}e^{-\im \omega t} G^\pm_\omega = \mp \im\Theta(\pm t)e^{\im (H+\im D)t},
\end{align}
Ref.~\cite{Cooper20} noted that $(G^+(t))_\eta = G^-(-t)$ is a
natural operation exchanging retarded and advanced propagators consistent with
causality. Combined with $\T$, this defines the operation $\T_\eta \equiv \T\circ \eta$ (cf. Eq.~(6a) of Ref.~\cite{Cooper20}),
there introduced as the ``unique way to extend Hamiltonian symmetries to Lindbladian
symmetries''. Our conclusions are different. As evidenced by
Eq.~\eqref{eq:CoviKeldyshExplicit}, the advanced and retarded propagators appear in
Lindbladian evolution in combination. Application of the standard antilinear
operation $\T$ (no $\eta$ involved) to the state exchanges $G^+ \leftrightarrow G^-$. Since these two operators appear in a paired configuration, cf. Eq.~\eqref{eq:CoviKeldyshExplicit}, the operation $\eta$ has no place in the symmetry analysis of Lindbladian state evolution. 

However,  it is of relevance in cases where the symmetries of the dissipation generators $K=H-\im D$ are considered in isolation.  In this case, $H=H_\T$ and $D=D_\T$ implies a symmetry $K_{\T_\eta}=H_\T -\im D_\T=K$, with physical consequences, e.g., in the statistical theory of the decay of resonances of open quantum systems (cf. Ref.~\cite{Fyodorov2003}). 

In view of the fact that a symmetry of $K$ under $T_\eta$ may have observable consequences for open quantum systems, one may ask just how general such symmetries are. For example, it is natural to expect that if a system \emph{and} its environment are time reversal invariant in the sense of unitary quantum mechanics, this symmetry is inherited by the dissipative operator $K$ after the integration over environmental degrees of freedom. In Appendix~\ref{app:anti} we show that this is not the case in general. Only in equilibrium does quantum mechanical time reversal invariance guarantee $T_\eta$ invariance of the reduced theory. This finding underpins our general statement that time inversion 
 out of equilibrium is meaningless in general. 
 
 Finally, we note that the classifications of non-Hermitian matrices $K = H - \im D$~\cite{Leclair02,Ueda19,Lee19} all have in common that they preserve the transformation laws of the Hermitian contribution $H$ under the anti-unitary symmetries familiar from the ground state classification. In light of the above discussion, such an extension to non-Hermitian matrices is only possible under conditions of global thermodynamic equilibrium.

\section{Conclusions}
\label{sec:conclusions}

In this paper, we classified the symmetries  governing the dynamics of open
fermionic quantum matter. Symmetries were defined as linear or anti-linear
transformations --- represented in Fock space, or in the first quantized language of
matrices for free systems --- leaving the irreversible equations of motion invariant. While this
rationale resembles the one  applied in the identification of symmetries in unitary
quantum time evolution, two principles make the out of equilibrium case different.
First, unitary state evolution, or, somewhat more generally, the micro-reversible
approach to a thermal equilibrium configuration, is governed by a single linear
operator, the system Hamiltonian. By contrast, the out of equilibrium generators
considered here comprise three linear operators, describing unitary evolution,
dissipation, and fluctuations, respectively. All three must obey individually defined
symmetry conditions for the full dynamics to be symmetric. The second difference
concerns time itself. In unitary dynamics, the application of anti-linear symmetries ($\im\to -\im$) is matched with an extraneously imposed inversion of time
($t\to -t$) to leave the quantum time evolution operator, $\exp(-\im\hat Ht)$, invariant. Within
the more general class of equilibrium processes, this operation is generalized to the
shift inversion $\E_\beta $, Eq.~\eqref{eq:EBetaDef}, 
($E_\beta f(t) = f(-t-i\beta)$),
likewise designed to keep the dynamics 
invariant. Combined with the anti-unitary Fock space transformation it defines the thermal time reversal $\T_\beta=\T \circ \E_\beta$, extending quantum mechanical time reversal to irreversible equilibrium dynamics. However,  out of equilibrium time reversal becomes
unphysical, meaning that anti-linear symmetries --- featuring in six out of ten
symmetry classes --- have a fundamentally different representation.

In view of these differences, it is remarkable that the stationary states stabilized
by equilibrium or non-equilibrium dynamics can be fully classified by \emph{identical} symmetries. For
example, when we say that a system is in symmetry class BDI, what we mean is that it
obeys an anti-linear $\T$-symmetry (squaring to unity) and a linear $\C$-symmetry,
likewise squaring to one. If the stationary state is Gaussian, its effective
Hamiltonian can be represented as a real matrix possessing a block-off diagonal
`chiral' structure. The identical realization of symmetries  relies on the stationarity of the asymptotic states --- where the meaning of time is lost, or, more formally, time inversion $\E_\beta$ acts as an identity and all symmetry operations reduce to their action in Fock space. It implies a strong principle of universality with obvious  practical consequences. Notably, the information contained in the periodic
table of topological insulators and superconductors universally applies to the
classification of Gaussian stationary states both in and out of equilibrium.

However, the equivalent representation of symmetries in the stationary limit does not
extend to the dynamical processes stabilizing them. Here, the presence or absence of $\E_\beta$ is key and leads to the identification of twenty `dynamical symmetry classes', distinct by the symmetry representations in either case. Ten of them describe the asymptotic approach towards a
stationary equilibrium configuration, the other ten the approach to stationary
non-equilibrium. Where the former assume the form of the ten well-known  symmetry
conditions for fermionic Hamiltonians, the latter require $30=3\times 10$ conditions
for the generators of unitary evolution, dissipation and fluctuations, respectively. The hallmark of the different symmetry representations in- and out of equilibrium is the sign difference in the $\T$-transformation of the Hamiltonian contribution to the generator of dynamics. 
For example, the approach to a BDI symmetric non-equilibrium stationary state, admits
a Hamiltonian contribution which, however, must be odd under the BDI symmetry rule
for Hermitian matrix generators, while in equilibrium evenness is required. Referring
back to the different treatment of time, the equilibrium symmetry classes reflect
invariance of the dynamical approach under the joint application of Fock space
symmetries and $\E_\beta$, while the non-equilibrium classes do not engage the
latter.

In this paper, we illustrated the above symmetry principles, and the consequences for  state
 topologies on the simple case of the BDI chain. However, it is relatively
 straightforward  to construct other
 realizations by reverse engineering. Starting from a model effective Hamiltonian $\Theta$ of specified symmetry
 and topological ground state, one thus asks which Lindblad operators $\hat L_\alpha$
 (see Eq.~\eqref{eq:lindblad}) stabilize this state. By design, the dissipation and
 fluctuation generators defining these operators via Eqs.~\eqref{eq:MatrixMDef} and
 ~\eqref{eq:DPDef} are then conditioned via the symmetry relations of
 table~\ref{tab2}. 

While the emphasis in this work has been on the bulk classification of phases, the next stage will be a more thorough exploration of the physics at the edge. Ideally, one would like to use the dissipatively stabilized topological edge as a resource for the storage and manipulation of quantum information. This requires the decoupling of the edge from the very physical mechanisms stabilizing it. For example, the edge space will be decoupled if it is realized as the dark space of an effective Lindblad equation. However, it will be interesting to explore if the isolation of the edge space can be effected in different ways, building on combined principles of symmetries and conservation laws as in Ref. \cite{Tonielli19}.

Another direction of research concerns the dynamical processes leading to a stationary limit. In this paper, stationarity was attained in a competition of dissipative damping ($K = H - \im D$), and fluctuations ($P$). However, there are alternative ways to describe the quantum stochastic process driving the relaxation: Inspection of the Lindblad equation shows that it contains the non-Hermitian combinations $H - P$ and $H+P$ as operators acting to the left and right of the density matrix, 
while $D - \im P$ acts  from both sides. In this decomposition, the first term describes the short time relaxation of quantum trajectories, interspersed by `quantum jumps' described by the last. The competition between the two can be accessed in dynamically resolved ways by post-selection or measurement protocols \cite{Lesanovsky10}. It will be interesting to explore topological signatures in full counting statistics \cite{Ren13,Riwar19} in the above language. This may define topological structures of the symmetry constrained matrix operators $H,D,P$ different from those considered in this paper.  

Finally, one may ask how bosonic systems fit into the general framework. While the
definition of fundamental symmetries extends to bosonic Fock spaces, the
manifestations of these symmetries in concrete states are strikingly different:
individual states can be multiply, or even macroscopically occupied, in which case
the dynamics becomes (semi)-classical and quantum noise less of an issue. The action
of state transformations confined by symmetries may be  non-compact
(think of the bosonic Bogoliubov transformation), and the stability of macroscopic
stationary states becomes an issue.
The latter compromises topology of Gaussian states of bosons \cite{Mink2019} and makes the presence of interactions
necessary \cite{Unanyan2020}.
 At the same time,  macroscopic bosonic quantum
states likely define a more natural application field for the physics of
non-Hermitian matrices than the strongly fluctuating fermion matter discussed here.

From a yet more general stance one may notice that, as with the physics of unitarily evolving quantum matter, the short range entangled symmetry protected fermionic phases considered here represent a relatively simple form of topological matter. With promising first steps taken in concerning the fate of fractional Quantum Hall states in open systems \cite{Yoshida2020}, the fascinating problem of extending the framework to dissipative variants of fractional or long range entangled matter is still out there and awaiting exploration. 

 \emph{Acknowledgements} -- We thank D. Borgnia, J. C. Budich, N. R. Cooper, V. Dwivedi, S. Lieu, F. Tonielli, M. McGinley, and M. Zirnbauer for discussions.  We acknowledge support from the Deutsche Forschungsgemeinschaft (DFG, German Research Foundation) under Germany's Excellence Strategy Cluster of Excellence Matter and Light for Quantum Computing (ML4Q) EXC 2004/1 390534769, and by the DFG Collaborative Research Centers (CRC) 183 Project No. 277101999 - project B02
 and 185 Project No. 277625399 - project C01. S.D. acknowledges support by the European Research Council (ERC) under the Horizon 2020 research and innovation program, Grant Agreement No. 647434 (DOQS). This research was supported in part by the National Science Foundation under Grant No. NSF PHY-1748958.
 
 \emph{We dedicate this work to the memory of our friend and colleague Federico Tonielli, whose brilliant promise in Theoretical Physics was cut short at age 27.}

\begin{appendix}

\section{Symmetry classes}
\label{sec:SymmetryClassesTable}
For the convenience of the reader, we here summarize the 10 symmetry classes, along with the dimensions where they admit topological ground states with $\Bbb{Z}$ or $\Bbb{Z}_2$ classification.
\begin{table}[h]
    \begin{tabular}{|c|ccc|cccc|}\hline
    Class&T&C&S&1&2&3&4\cr\hline
    A&0&0&0&0&$\Bbb{Z}$&0&$\Bbb{Z}$\cr
    AIII&0&0&+&$\Bbb{Z}$&0&$\Bbb{Z}$&0\cr\hline
    AI&+&0&0&0&0&0&$\Bbb{Z}$\cr
    BDI&+&+&+&$\Bbb{Z}$&0&0&0\cr
    D&0&+&0&$\Bbb{Z}_2$&$\Bbb{Z}$&0&0\cr
    DIII&$-$&+&+&$\Bbb{Z}_2$&$\Bbb{Z}_2$&$\Bbb{Z}$&0\cr
    AII&$-$&0&0&0&$\Bbb{Z}_2$&$\Bbb{Z}_2$&$\Bbb{Z}$\cr
    CII&$-$&$-$&+&$\Bbb{Z}$&0&$\Bbb{Z}_2$&$\Bbb{Z}_2$\cr
    C&0&$-$&0&0&$\Bbb{Z}$&0&$\Bbb{Z}_2$\cr
    CI&+&$-$&+&0&0&$\Bbb{Z}$&0\cr\hline
    \end{tabular}
    \caption{Periodic table of topological insulators. The first two rows contain the classes without anti-untiary symmetries, for brevity we use the labels $\pm\equiv \pm 1$.\label{tab:PeriodicTable}}
\end{table}
The labels $\Bbb{Z},\Bbb{Z}_2,0$ in the table denote the possible topological invariants. For
example, the system considered in section~\ref{sec:case_study} is defined in $d=1$ and symmetry class
$(\T,\C,\SC)=(+1,+1,1)$, or $\mathrm{BDI}$.

\section{Action of unitary symmetries}
\label{app:unitaryex}


Here we elaborate on the condition, that anti-unitary symmetries $\T,\C,\SC$ be realized within the irreducible representation spaces of the system's unitary symmetries $\U$. This is best explained on the basis of examples. 
 For instance, the plain operator exchange  $\C a_i \C^{-1}=a_i^\dagger$ defines  a symmetry structure of every free fermion operator, $O$,
  entirely on the basis of Fermi statistics: the operation  acts on the Nambu operators as
 $A_i\to A_i^\dagger= \sigma_x A_i$, and using the equivalence of these
 representations, $A_i=\sigma_x A_i^\dagger$, we obtain
 \begin{align}
 \label{Eq:FermiSymmetry}
   A^{\dagger T} O A = A^T \sigma_x O \sigma_x A^\dagger=-A^{\dagger T} \sigma_x O^T \sigma_x A+\mathrm{tr}(O),
 \end{align}
 where the trace comes from the Kronecker $\delta$ in $\{ a^\fd_i,a^\dag_j\} =\delta_{ij}$.
 Momentarily ignoring the trace, we have the relation $O=-\sigma_x O^T \sigma_x$,
 which for Hermitian  $O$ defines an operator of class D. Now
 consider a situation with particle number conservation. In this case, $\hat O$ commutes
 with the unitary operator  $\U=\exp(\im\alpha \hat N)$, where $\hat N=\sum a^\dagger_i a^\fd_i$ and $\alpha \in \mathfrak{u}(1)$.
 However, $\C$ does not, $\C \hat N \C^{-1}=N-\hat N$. The operation $\C$ couples different sectors of conserved particle number such that the above principle is violated. To understand the consequences, note that  
 number conservation implies block diagonality in Nambu space, $O\equiv \mathrm{bdiag}(o,-o^T)$. The operation $\C$ connects the two blocks, however, absent a
 physical coupling remains meaningless. The example illustrates how a second quantized
 operator of relatively higher symmetry (number conservation) can have a lesser
 symmetry on the matrix level (just Hermiticity, class A, rather than D)  --- in other words, symmetries or `structures' on the first quantized level need not be rooted in actual symmetries of the many-body context. 
 
 However, if particle number conservation is violated by, e.g., an `order parameter' $a_i^\dagger
 \Delta_{ij} a_j^\dagger$, the previously isolated sectors of definite number get combined to an enlarged representation space. $\C$ now acts \emph{within} this space and does define a meaningful `BCS' matrix structure
 $O=\left(\begin{smallmatrix}o&\Delta\cr \Delta^\dagger &-o^T
 \end{smallmatrix}   \right)$ with class D symmetry $O=-\sigma_x O^T\sigma_x$.
 Similarly, consider the example of spin rotation symmetry from Sec. \ref{sub:symmetries}, $\U_s$, where  the
 plain $\C:a_i \rightarrow a_i^\dagger$  does not commute and violates the unitarity
 principle,  while $\C_s: a_{\sigma}
 \rightarrow(\sigma_y)_{\sigma\sigma'}a^\dagger_{\sigma'}$ does not. The operation
 $\C_s$ thus defines a symmetry class (C) as $O= -\sigma_y O^T \sigma_y$.


\section{Symmetries in Gaussian state evolution} 
\label{sec:appendix_symmetries_of_gaussian_states}

We here derive table \ref{tab2} 
by subjecting the Lindblad equation~\eqref{eq:LindbladianDPH} to the symmetry operations in their second quantized incarnation, Eq.~\eqref{eq:UTCSecondQuantized}. 

\noindent \emph{$\T$-invariance --} Application of  $\T$ in second quantized incarnation leads to 
\begin{align*}
  \partial_t\hat \rho_\T=&\, \im (\hat H_\T -\hat P_\T)\hat \rho_\T - \im \hat \rho_\T (\hat H_\T + \hat P_T) \cr 
   &+ 2 A^T\tfrac{1}{2}(\bar D_\T -\im\bar P_\T)\hat \rho_\T A^\dagger- c\hat \rho_\T.
\end{align*}
This equation becomes identical to the untransformed one, provided $\hat H_\T=-\hat H, \hat P_\T=-\hat P$, and $\bar D_\T=\bar D$, $\bar P_\T=-\bar P$. Comparison with table~\ref{tab2} shows that these conditions are met if the matrices $H,P,D$ satisfy the conditions listed in the first row.
\smallskip

\noindent \emph{$\C$-invariance --} In a similar manner, the application of $\C$ yields
\begin{align*}
  \partial_t\hat \rho_\C &= - \im (\hat H_\C -\hat P_\C)\hat \rho_\C + \im \hat \rho_\C (\hat H_\C + \hat P_C) \cr 
   &\quad\,\,+2A^{\dagger T} U_\C^T\tfrac{1}{2} (\bar D +\im\bar P)\hat \rho_\C \bar U_C A- c\hat \rho_\C.
\end{align*}
We change the representation of the jump term in the second line as
\begin{align*}
& A^{\dagger T} U_\C^T (\bar D +\im\bar P)\hat \rho_\C \bar U_C A\stackrel{\eqref{eq:USigmaX}}=A^{T} U_\C^\dagger  \sigma_x(\bar  D +\im\bar P)\sigma_x U_\C\hat \rho_\C   A^\dagger\cr 
  &  \stackrel{(\ref{eq:prop1},\ref{eq:prop2})}{=}A^{T} U_\C^\dagger  (D +\im P) U_\C\hat \rho_\C   A^\dagger \stackrel{\eqref{eq:UTCFirstQuantized}}{=}
   A^{T}   (-\bar{D}_\C +\im\bar{P}_\C) \hat \rho_\C   A^\dagger,
\end{align*}
and substitution back into the equation yields
\begin{align*}
   \partial_t\hat \rho_\C=&\, -\im(\hat H_\C -\hat P_\C)\hat \rho_\C + \im\hat \rho_\C (\hat H_\C + \hat P_C)\cr 
   &+ 2A^{T}  \tfrac{1}{2} (-\bar{D}_\C +\im\bar{P}_\C) \hat \rho_\C   A^\dagger- c\hat \rho_\C.
\end{align*}
By the same rationale as in the previous case, the invariance of the equation requires the matrix transformations listed in the second row of table~\ref{tab2}.
\smallskip

\noindent \emph{$\SC$-invariance --} Testing for $\SC$ is not necessary, as it is a consequence of the combined presence of $\C$ and $\T$. However, it is instructive to see how this symmetry manifests itself in the Lindblad equation without reference to the composition. Application of $\SC$ leads to
\begin{align*}
  \partial_t\hat \rho_\SC=&\,\im(\hat H_\SC -\hat P_\SC)\hat \rho_\SC - \im \hat \rho_\SC (\hat H_\SC + \hat P_T) \cr 
   &+2A^{\dagger T} U_\SC^T \tfrac{1}{2}( D -\im P)\hat \rho_\SC \bar U_\SC A- c\hat \rho_\SC. 
\end{align*}
Once again, the term in the second line requires special attention:
\begin{align*}
  & A^{\dagger T} U_\SC^T ( D -\im P)\hat \rho_\SC \bar U_\SC A\stackrel{\eqref{eq:USigmaX}}=A^{T} U_\SC^\dagger  \sigma_x(  D -\im P)\sigma_x U_\SC\hat \rho_\SC   A^\dagger\cr 
  &  \stackrel{(\ref{eq:prop1},\ref{eq:prop2})}{=}A^{T} U_\SC^\dagger  (\bar D -\im\bar P) U_\SC\hat \rho_\SC   A^\dagger \stackrel{\eqref{eq:UTCFirstQuantized}}{=} A^{T}   (-{D}_\SC -\im{P}_\SC) \hat \rho_\SC   A^\dagger.
\end{align*}
Substitution into the equation leads to 
\begin{align*}
  \partial_t\hat \rho_\SC=&\,\im(\hat H_\SC -\hat P_\SC)\hat \rho_\SC - \im\hat \rho_\SC (\hat H_\SC + \hat P_T) \cr 
   &+2A^{T}   \tfrac{1}{2}(-\bar{D}_\SC -\im\bar{P}_\SC) \hat \rho_\SC   A^\dagger- c\hat \rho_\SC,
\end{align*}
and upon comparison with the untransformed equation to $\hat H_\SC=-\hat H$, $\hat P_\SC=-\hat P$, $\hat D_\SC=- \hat D$, or the third line in table \ref{tab2} as a condition for invariance $\hat \rho=\hat \rho_\SC$.






\section{Symmetries in Keldysh representation} 
\label{sec:symmetries_in_keldysh_representation}

We here discuss how the symmetry of Gaussian states as expressed by the last column of table~\ref{tab2} leads to Eqs. \eqref{eq:SymmetriesKeldyshKernel}  for the generally non-Markovian matrix kernels $\mathcal{M}$ generating the dynamics.
\smallskip

\noindent \emph{$\T$-invariance --} According to table \ref{tab2}, $\T$-invariance means the existence of a unitary matrix $U_\T$ such that $\Gamma=\Gamma_
\T=U_\T \bar\Gamma U_\T^\dagger$. Starting from the representation~\eqref{eq:KeldyshFunctional} this becomes
\begin{align}
\label{eq:CoviTKeldysh}
 & (U_\T \overline \Gamma U_\T^\dagger)_{ab}  \\
 &=U_{\T aa'}\int  e^{-\tfrac{\im}{2}\int (d\omega)\, \eta_{\omega}^T(\omega \tau_1 -\overline{\M_\omega})\nu_{-\omega}} \nu^c_{a'}(t)\eta^c_{b'}(t)U^\dagger_{\T b'b}\cr
 &=\int  e^{\tfrac{\im}{2}\int (d\omega)\, \eta^T_{-\omega}(\omega \tau_1 +U_\T\overline{\M_{-\omega}}U_\T^\dagger) \nu_{\omega}}\, \nu^c_a(t)\eta^c_{b}(t)\stackrel{!}{=}\Gamma_{ab}(t).\nonumber
\end{align}
In the third line, we transformed variables $U_\T \nu\rightarrow \nu$ and $\eta^T U_\T^\dagger \rightarrow \eta^T$ (note that $ U_\T \omega \tau_1 U_\T^\dagger = \omega \tau_1$), and changed $\omega\to -\omega$  in the frequency integral. A sufficient condition for the invariance of the $\Gamma$ matrix then is $ (\mathcal{M}_\T)_\omega= U_\T\overline{\M_{-\omega}} U_\T^\dagger=-\mathcal{M}_\omega$, which is the first line of Eq.~\eqref{eq:SymmetriesKeldyshKernel}.
\smallskip

\noindent \emph{$\C$-invariance --} In the path integral formalism,  $\C$-invariance, $\Gamma=\Gamma_\C=-U^T_\C \Gamma^T \bar U_\C$, is probed as
\begin{align*}
 & (-U^T_\C  \Gamma^T \bar U_\C)_{ab}  =
 - U^\dagger_{\C bb'}\Gamma^\fd_{b'a'}U^\fd_{\C a'a} \cr 
 &=-U^\dagger_{\C bb'}\int  e^{\tfrac{\im}{2}\int (d\omega) \,\eta^T_{-\omega}(\omega \tau_1 -\M_\omega)\nu_\omega}\, \nu^c_{b'}(t)\eta^c_{a'}(t)U_{\C a'a}\cr 
 &=-\int  e^{\tfrac{\im}{2}\int (d\omega)\, \eta^T_{-\omega} U^\dagger_\C(\omega \tau_1-\M_\omega)
 U_\C \nu_\omega} \nu^c_{b}(t)\eta^c_{a}(t)\cr 
 &=\int  e^{\tfrac{\im}{2}\int (d\omega)\,  \nu^T_{-\omega} (\omega \tau_1+U^T_\C\mathcal{M}_{-\omega}^T\bar U_\C)
  \eta_{\omega}^T } \eta^c_{a}(t) \nu^c_{b}(t)\stackrel{!}{=}\Gamma_{ab}(t),
\end{align*}
where in the crucial fourth line we swapped the order of variables both in the action and the pre-exponential variables (picking up a sign in the process). Except for a different naming of the dummy variables, $\eta \leftrightarrow \nu$, the final expression equals the original one, provided $\mathcal{M}$ satisfies the $\C$ entry in Eq.~\eqref{eq:SymmetriesKeldyshKernel}.
\smallskip

\noindent \emph{$\SC$-invariance --} Finally, $\SC$-invariance, $\Gamma=\Gamma_\SC=-U_\SC^T\Gamma^\dagger \bar U_\SC$ is established as
\begin{align*}
 & (-U_\SC^T  \Gamma^\dagger \bar U^\fd_\SC)_{ab}  =
 - U^\dagger_{\SC bb'}\bar \Gamma^\fd_{b'a'}U_{\SC a'a} \cr 
 &=-U^\dagger _{\SC bb'}\int  e^{-\tfrac{\im}{2}\int (d\omega)\, \eta^T_{\omega}(\omega \tau_1 -\overline{\M_\omega})\nu_{-\omega}} \,\nu^c_{b'}(t)\eta^c_{a'}(t) U^\fd_{\SC a'a}\cr 
 &=-\int  e^{-\tfrac{\im}{2}\int (d\omega)\, \eta^T_{\omega} ( \omega \tau_1-U^\dagger_\SC\overline{\M_\omega}U_\SC)
  \nu_{-\omega}}\, \nu^c_{b}(t)\eta^c_{a}(t)\cr 
  &=\int  e^{\tfrac{\im}{2}\int (d\omega)\,  \nu^T_{-\omega} (\omega \tau_1-U_\SC^T (\M_\omega)^\dagger \bar U^\fd_\SC)
  \eta^T_{\omega}}\, \eta^c_{a}(t) \nu^c_{b}(t)\stackrel{!}{=}\Gamma_{ab}(t),
 \end{align*}
 which leads to the final entry in  Eq.~\eqref{eq:SymmetriesKeldyshKernel}.

\section{Proof of Eqs.~\eqref{eq:TDaggerTrace} and \eqref{eq:KMSCombinedWithT}} 
\label{sec:proof_of_eq_eq:tdaggertrace}

Consider a general second quantized $q$-body operator with non-vanishing trace: $\hat
Y=\sum'_{i,j}Y_{i_1,\dots,i_q,j_q,\dots,j_1}a^\dagger_{i_1}\dots
a^\dagger_{i_q}a^\fd_{j_q}\dots a^\fd_{j_1}$, where the coefficients
$Y_{i_1,\dots,i_q,j_q,\dots,j_1}$ are anti-symmetric under pairwise exchange of $i$-
and $j$-indices among themselves (Fermi statistics), and the primed sum $\sum'$
extends over ordered indices $i_1<\dots<i_q$, $j_1<\dots<j_q$.  The trace of this
operator is readily obtained as $\mathrm{tr}(\hat Y)=\sum'_{i}
Y_{i_1,\dots,i_q,i_q,\dots,i_1} \mathrm{tr}(\hat n_{i_1}\dots \hat
n_{i_q})=2^{N-q}\sum'_{i}
Y_{i_1,\dots,i_q,i_q,\dots,i_1}$, where $N$ is the dimension of the single particle Hilbert space, and  $\hat n_i=a^\dagger_i a^\fd_i$, or
\begin{align}
\label{eq:AppYTrace}
 \mathrm{tr}(\hat Y)= \frac{2^{N-q}}{q!}\sum_iY_{i_1,\dots,i_q,i_q,\dots,i_1},
\end{align}
with an  unrestricted index summation.

With $\hat Y^\dagger=\sum'_{i,j}\bar Y_{i_1,\dots,i_q,j_q,\dots,j_1}a^\dagger_{j_1}\dots
a^\dagger_{j_q}a^\fd_{i_q}\dots a^\fd_{i_1}$, the same construction yields
\begin{align*}
 \mathrm{tr}(\hat Y^\dagger)=\frac{2^{N-q}}{q!}\sum_i\bar Y_{i_1,\dots,i_q,i_q,\dots,i_1}. 
\end{align*}
Now consider $\hat Y_\T=\sum'_{i,j} Y_{\T i_1,\dots,i_q,j_q,\dots,j_1}   a^\dagger_{i_1}  \dots
a^\dagger_{i_q}a^\fd_{j_q}\dots a^\fd_{j_1}$, where the transformed coefficients $Y_{\T i_1,\dots,i_q,j_q,\dots,j_1}\equiv \bar Y_{i'_1,\dots,i'_q,j'_q,\dots,j'_1} \bar U_{\T i'_1 i_1}\dots \bar U_{\T i'_q i_q} U_{\T j'_q j_q}\dots \bar U_{\T j'_1 j_1}$. The trace is obtained via Eq.~\eqref{eq:AppYTrace} as 
\begin{align*}
  \mathrm{tr}(\hat Y_\T)&= \frac{2^{N-q}}{q!}\sum_i Y_{
  \T i_1,\dots,i_q,i_q,\dots,i_1}\cr 
  &= \frac{2^{N-q}}{q!}\sum_i \bar Y_{
   i_1,\dots,i_q,i_q,\dots,i_1}=\mathrm{tr}(\hat Y^\dagger), 
 \end{align*} 
where in the second equality the unitarity of the matrices $U_\T$ was used. With $\hat {\Theta}=\hat Y^\dagger$ we arrive at Eq.~\eqref{eq:TDaggerTrace}.

To prove Eq.~\eqref{eq:KMSCombinedWithT}, we apply Eq.~\eqref{eq:TDaggerTrace} to the operator $\hat {\Theta}\equiv \hat \rho\hat B\hat A_{t+\im\beta}$ in Eq.~\eqref{eq:KMSRelation}. Assuming hermiticity of $\hat B$ and $\hat A$, we obtain $\hat {\Theta}^\dagger = \hat A_{t-\im\beta} \hat B \hat \rho$, and the subsequent $\T$-operation gives
$\hat {\Theta}^\dagger_\T= \T \hat {\Theta}^\dagger \T^{-1}= (\T \hat A_{t-\im\beta}  \T^{-1})(\T \hat B \hat  \T^{-1}) (\T \hat \rho\T^{-1})=  (\hat A_\T)_{-t-\im\beta}  \hat B_\T \hat \rho$, where $\hat H=\hat H_\T$ was used. Substitution into Eq.~\eqref{eq:KMSRelation} yields Eq.~\eqref{eq:KMSCombinedWithT}. 

\section{Proof of Eq.~\eqref{eq:HDPEquilibrium}} 
  \label{sec:proof_of_eq_xx}

We here prove how the invariance condition~\eqref{eq:HDPEquilibrium} follows if the theory is symmetric under the combined application of $\T_\beta = \T \circ \E_\beta$ to the Keldysh functional. We will see that the symmetry condition in turn relies on the fluctuation-dissipation relation Eq.~\eqref{eq:FDT}. For concreteness and the sake of easy comparability to the previous discussion, we monitor the consequences of  the symmetry for the stationary long time limit of the covariance matrix, 
\begin{align*}
  \Gamma_{ab} \equiv \int \, e^{\tfrac{\im}{2}\int (d\omega)\, \eta^T_{-\omega}( \omega \tau_1 -\M_\omega)\nu_\omega}\, \int_{-\infty}^\infty (d\omega) \, \nu^c_{a,\omega} \eta^c_{b,-\omega}.
\end{align*}
Proceeding as in Eq.~\eqref{eq:CoviTKeldysh}, we apply the $\T$-symmetry to obtain
\begin{align*}
  &(U_\T \bar \Gamma U_\T^\dagger)_{ab} =\cr 
  &\quad =    \int \, e^{-\tfrac{\im}{2}\int (d\omega)\, \eta^T_{\omega}( \omega \tau_1 -\M_{-\omega,\T})\nu_{-\omega}}\, \int (d\omega) \, \nu^c_{a,-\omega} \eta^c_{b,\omega},
\end{align*}
where $\M_{-\omega,\T}=U_\T \overline{\M_{\omega}}U^\dagger_\T$ (cf. Eq.~\eqref{eq:SymmetriesKeldyshKernel}).
We now take one more step to apply the  $\E_\beta$ transformation, which is represented on the Keldysh
integration variables through Eq.~\eqref{eq:EbetaKeldyshFields}. A first observation is that this transformation leaves the preexponential terms invariant. (To see this, rearrange the latter as $\nu_a^c
\eta_b^c \rightarrow - \eta_b^c \nu_a^c \rightarrow - \eta_b^c \nu_a^c  - \eta_b^q
\nu_a^q =- \eta^T_b \nu_a$, where in the final second step we noted that expectation
values of purely quantum type, $\langle \eta_b^q \nu_a^q\rangle $, vanish in a 
Keldysh theory. In this representation, the transformation Eq.~\eqref{eq:EbetaKeldyshFields} drops out due to $\mathcal{Y}^T_{-\omega} \mathcal{Y}_\omega=\mathds{1}$.) Then, 
\begin{align*}
  &(U_\T \bar \Gamma U_\T^\dagger)_{ab} =\cr 
  &\quad =    \int \, e^{\tfrac{\im}{2}\int (d\omega)\, \eta^T_{-\omega}( \omega \tau_1 +\mathcal{Y}^T_{\omega}\M_{\omega,\T}\mathcal{Y}_{-\omega})\nu_{\omega}}\, \int (d\omega) \, \nu_{a,\omega}^c\eta^c_{b,-\omega}.
\end{align*}
Comparing with the original representation, we find that $U_\T \bar \Gamma U_\T^\dagger=\Gamma$ if
\begin{align}
  \mathcal{Y}^T_{\omega}\M_{\omega,\T}\mathcal{Y}_{-\omega}=-\M_\omega.
\end{align}
We now need to investigate what this relation implies for the matrix blocks defining the Kelysh operator $\mathcal{M}$ through Eq.~\eqref{eq:KeldyshKernelMDef}. This is best done in a Pauli matrix decomposition,
\begin{widetext}
\begin{align*}
  \mathcal{M}_\omega&=H_\omega \tau_x - D_\omega \tau_y - P_\omega (\mathds{1}-\tau_z),\cr
   \mathcal{M}_{\omega,\T}&=H_{\omega,\T} \tau_x + D_{\omega,\T} \tau_y - P_{\omega,T} (\mathds{1}-\tau_z),\cr
   \mathcal{Y}^T_{\omega}\M_{\omega,\T}\mathcal{Y}_{-\omega}&=  
   -H_{\omega,\T} \tau_x + D_{\omega,\T} e^{\omega \beta \tau_x}\tau_y - P_{\omega,T} (\mathds{1}+ e^{\beta \omega \tau_x}\tau_z)=\cr 
   &=-H_{\omega,\T} \tau_x + D_{\omega,\T}(\ch\, \tau_y +i \sh \tau_z)-P_{\omega,T} (\mathds{1}+ \ch \tau_z -i \sh \tau_y).
\end{align*}
\end{widetext}
where we used the abbreviations $\ch=\cosh(\beta\omega)$ and $\sh=\sinh(\beta\omega)$.  Comparison of the linearly independent contributions multiplying $\tau_{y},\tau_z,\mathds{1}$ then readily leads to Eq.~\eqref{eq:HDPEquilibrium} and the constraint Eq.~\eqref{eq:FDT}.
\section{Alternative approach to symmetry classification?} 
  \label{app:anti}

  Our approach to symmetry classification is based on the idea that equilibrium
  dynamics  -- and even more restrictedly, unitary Hamilton dynamics -- should be
  viewed as a special case of more general non-equilibrium evolutions. However, one
  may also approach the situation from an opposite perspective. Its starting point is
  a microscopic Hamiltonian $\hat H_\text{t}\equiv \hat H_\text{s}+\hat H_\text{b}+\hat H_\text{c} $ describing a system ($\hat H_\text{s}$),  an environment ($\hat H_\text{b}$), and their coupling ($\hat H_\text{c}$). Elimination of the environment makes the system dynamics irreversible.
  Assuming that symmetries are preserved in the process, one may attempt a classification based on the symmetries of the
  microscopic Hamiltonian. For example,  a time reversal invariant microscopic
  Hamiltonian, $\hat H_\text{t}=\hat H_{\text{t},\T}$ would then define a system dynamics  with inherited $\T$-invariance.
  Specifically,  the matrix generator $K=H -\im D$ containing the  quadratic contribution of the system  Hamiltonian in first quantized language, $H$, and the damping due to environmental coupling would satisfy $H= H_{\T}$ and $D=D_\T$ and in the consequence  invariance under $\T_\eta\equiv \T\circ \eta$: $H-\im D =
  (H-\im D)^\dagger_\T=H_T - \im D_\T$. 

  However, there is a loophole in this argument. It ignores the \emph{state} of the environment, as described by its distribution functions. (Unlike the transient state of the system, which does not play a role in the symmetry classification, the states of the environment are robust as per definition of the term `environment'.) In the following, we consider a case study illustrating how the violation of $\T$-invariance by a nonequilibrium environmental distribution can break the invariance of the 
  effective system dynamics, $K\not= K_\T$.

\textit{Microscopic model} -- We consider a model with two fermionic modes $a = (a_1,a_2)$, each coupled to a bath with modes $b_\mu = (b_{\mu,1} ,b_{\mu, 2})$. We choose bath temperatures $T_{1,2}$, which for $T_1\not=T_2$ breaks equilibrium. The system-bath Hamiltonian reads
\begin{align}\label{eq:hsb}
\hat H_{\text{t}} &= \hat H + \hat H_{\text{int}}+ \hat H_\text{c}  + \hat H_{\text{b}} ,\\
& \hat H = a^\dag (h_x \Sigma_x  + h_y \Sigma_y) a, \qquad  \hat H_\text{int} = \lambda  a_1^\dag a_1  a_2^\dag a_2,\cr 
&  \hat H_{\text{c}} = \sum_{\mu} g_\mu[ a^\dag  {\mathbb{1}} b_{\mu} +  b_{\mu}^\dag  {\mathbb{1}} a ] ,\quad  \hat H_{\text{b}} = \sum_{\mu}    b_{\mu}^\dag \epsilon_\mu {\mathbb{1}} b_{\mu}.\nonumber 
\end{align} 
Here, the $\mathbb{1}, \Sigma_{x,y,z}$ act in band space and $\hat H_\text{s}\equiv \hat H + \hat H_\text{int}$ describes the system Hamiltonian, featuring a free part and a  density-density self-interaction, which we will see plays an important role. The terms $\hat H_{\text{c}}$ and $ \hat H_{\text{b}} $ model the coupling to a harmonic bath. With the real couplings, $g_\mu$,  the above system-bath Hamiltonian is time reversal invariant, with transformations in band space as
\begin{eqnarray}\label{eq:trtr}
U^a_\T =\Sigma_x,\quad U^{b_\mu} _\T =  \Sigma_x \quad\,\forall \mu .
\end{eqnarray}
The microscopic Keldysh action for the non-interacting contribution to this setting reads, in the frequency domain (from now on, $a = (a_{c,1},a_{c,2},a_{q,1},a_{q,2})^T, b_\mu = (b_{c,\mu_1},b_{c,\mu_2},b_{q,\mu_1},b_{q,\mu_2})^T$)
\begin{widetext}
\begin{eqnarray}
S_0 &=& \int(d\omega) \bar a^T (\omega -  H )\tau_x  a , \quad S_{\text{c}} =  \int (d\omega) \sum_{\mu} g_\mu[  \bar a^T {\mathbb{1}} \otimes \tau_x b_{\mu} +  \bar b^T_{\mu}  {\mathbb{1}}\otimes\tau_x  a] , \\\nonumber
 S_{\text{b}} &=& \int (d\omega) \sum_{\mu} \bar b^T_{\mu} G_{\mu,\omega}^{-1} b_{\mu}, 
\quad
 G^{-1}_{\mu,\omega} = \left(\begin{array}{cc}
0   &( \omega - (\epsilon_{\mu} + \im\kappa)){\mathbb{1}}  \\
(\omega - (\epsilon_\mu - \im\kappa)){\mathbb{1} } &   2 \im \kappa ( t^+_\omega  {\mathbb{1}}+ t^-_\omega  \Sigma_z   ) 
\end{array}\right),
\end{eqnarray}
\end{widetext}
where $H$ is the above system Hamiltonian matrix, the Pauli matrix $\tau_x$ acts in Keldysh space, and $t^\pm_\omega = \tfrac{1}{2}(\tanh \tfrac{\beta_1 \omega }{2}\pm \tanh \tfrac{\beta_2 \omega}{2} )$. The infinitesimal parameter $\kappa >0$ defines the causality of the Green functions, and it fixes the state of the bath: inverting the above matrix we find $G_{cc,\mu}=2\pi \im \delta(\omega-\epsilon_\mu)( t^+_\omega  {\mathbb{1}}+ t^-_\omega  \Sigma_z   )$, where the departure from equilibrium is measured by a non-zero coefficient $t^-_\omega $, and the appearance of a matrix $\Sigma_z$ in band space. Notice that the distribution mismatch breaks the $\T$-invariance of the action, $\Sigma_{z,\T}=-\Sigma_z$. Although the distribution functions couple to the action only infinitesimally via $\kappa$, they do feed back into the system dynamics on an  $\mathcal{O}(1)$ level, as the following discussion shows.

\textit{Effective Lindblad model} -- Integrating out the bath, we obtain the effective system action
\begin{align}\label{eq:seff}
&S_\text{eff} = \int(d\omega) \bar a^T G^{-1}_{\text{eff},\omega} a - \lambda  \int dt ( n_{1c}n_{2q} +n_{1c}n_{2q}),
\end{align}
with
\begin{align*}
 G^{-1}_{\text{eff},\omega}& =  \left(\begin{array}{cc}
0   &  (G^{-}_{\text{eff},\omega})^{-1} \\
(G^{+}_{\text{eff},\omega} )^{-1}&   2 P_{\text{eff},\omega}
\end{array}\right) =  \cr 
&=\left(\begin{array}{cc}
0   & \omega{\mathbb{1}} - (H + \im d_\omega {\mathbb{1}}) \\
\omega {\mathbb{1}} - (H- \im d_\omega{\mathbb{1}}) &   2 \im d_\omega (t^+_\omega  {\mathbb{1}}+ t^-_\omega  \Sigma_z   ) 
\end{array}\right),
\end{align*}
where $d_\omega  = \pi \sum_\mu  g^2_\mu \delta(\epsilon_\mu - \omega)$ and $n_{ic} = \bar a_{ic} a_{ic}+ \bar a_{iq} a_{iq} , n_{iq} = \bar a_{ic} a_{iq}+ \bar a_{iq} a_{ic}$, $i=1,2$, and we neglected a Lamb shift renormalizing the system Hamiltonian, which is unimportant for the present discussion.

At this level, it looks like the induced matrix generator $K \equiv H - \im  d_\omega \mathds{1}$ is $\T$-symmetric. However, this changes once the interaction Hamiltonian is taken into account. To first order in perturbation theory, this generates a self-energy correction $\sim
\lambda \int (d\omega) G^+_{\text{eff},\omega}
P_{\text{eff},\omega}G^-_{\text{eff},\omega}$, where the $\Sigma_z$ matrix contained in
$P_{\text{eff} , \omega}$ reflects the absence of equilibrium. Substituting this expression into Eq.
\eqref{eq:seff}, we induce a term  $\propto  \Sigma_z$ in $K$. This term is a consequence of the sensitivity of the self energy to the bath distribution functions, which for $T_1\not= T_2$ break time reversal. In this case, $K_\T\not=K$ and the symmetry under non-Hermitian time reversal $\T_\eta$~\cite{Cooper20} is lost.  

The upshot of the above discussion is that out of equilibrium, the microscopic $\T$-symmetry of a system+environment Hamiltonian does not stabilize an induced $T_\eta$ symmetry of the matrix generator $K$. Only in equilibrium, the full symmetry of the theory (including the bath distribution functions) under $\T_\beta$ descends to a  $T_\eta$ symmetry of this operator, as outlined in section~\ref{sec:beyond_the_Markovian_limit}. Out of equilibrium, the only way to stabilize an antilinear $\T$-symmetry of a dynamical evolution and its stationary phases (then without reference to time reversal) is via generators defining the criteria of table~\ref{tab2}.

\end{appendix}

\bibliography{bibliography}


\end{document}